\documentclass[aps,prx,twocolumn,amsmath,amssymb,superscriptaddress,showpacs,nobalancelastpage,floatfix,longbibliography]{revtex4-1}
\usepackage{amsmath,amssymb,physics}
\usepackage{graphicx}
\usepackage{appendix}
\usepackage[usenames,dvipsnames]{color}
\usepackage[colorlinks=true, citecolor=blue, linkcolor=blue, urlcolor=blue]{hyperref}
\usepackage{bm}
\usepackage{bbm}
\newcommand{\jl}[1]{\textcolor{black}{#1}}

\usepackage[nottoc,numbib]{tocbibind}

\begin{document}

\title{Quantum to classical crossover of Floquet engineering  in correlated quantum systems}


\author{Michael A.~Sentef}
\email{michael.sentef@mpsd.mpg.de}
\affiliation 
{Max Planck Institute for the Structure and Dynamics of Matter, Luruper Chaussee 149, 22761 Hamburg, Germany}

\author{Jiajun Li}
\affiliation 
{Department of Physics, University of Erlangen-Nuremberg, 91058 Erlangen, Germany}

\author{Fabian K\"unzel}
\affiliation 
{Department of Physics, University of Erlangen-Nuremberg, 91058 Erlangen, Germany}

\author{Martin Eckstein}
\email{martin.eckstein@fau.de}
\affiliation 
{Department of Physics, University of Erlangen-Nuremberg, 91058 Erlangen, Germany}

\date{\today}

\begin{abstract}
Light-matter coupling involving classical and quantum light offers a wide range of possibilities to tune the electronic properties of correlated quantum materials. Two paradigmatic results are the dynamical localization of electrons and the ultrafast control of spin dynamics, which have been discussed within classical Floquet engineering and in the deep quantum regime where vacuum fluctuations modify the properties of materials. Here we discuss how these two extreme limits are interpolated by a cavity which is driven to the excited states. In particular, this is achieved by formulating a Schrieffer-Wolff transformation for the cavity-coupled system, which is mathematically analogous to its Floquet counterpart. 
Some of the extraordinary results of Floquet-engineering, such as the sign reversal of the exchange interaction or electronic tunneling, which are not obtained by coupling to a dark cavity, can already be realized with a single-photon state (no coherent states are needed). The analytic results are verified and extended with numerical simulations on a two-site Hubbard model coupled to a driven cavity mode. Our results generalize the well-established Floquet-engineering of correlated electrons to the regime of quantum light. It opens up a new pathway of controlling properties of quantum materials with high tunability and low energy dissipation.
\end{abstract}
\maketitle

\section{Introduction}
\label{Sec:introduction}
Under a time-periodic perturbation, such as the electric field of a laser or a coherently excited phonon, the time-evolution and steady states of a quantum system are described  by an effective time-independent Floquet Hamiltonian $H^F$, which can be entirely different from the undriven one. Mathematically, $H^F$ is defined through the stroboscopic time-evolution $U(t+T,t)\equiv\exp(-iTH^F)$ over a period $T=2\pi/\Omega$ of the drive. The design of a given Floquet Hamiltonian with suitable driving protocols, termed Floquet engineering \cite{Bukov2015,Eckardt2017,oka_floquet_2019}, has become an important tool for quantum simulation with ultracold gases, and it has been widely discussed in relation to the control of interactions and phase transitions in solids. A certainly incomplete list of theoretical proposals includes the manipulation of topologically nontrivial bands \cite{Oka2009,Lindner2011}, spin Hamiltonians \cite{mentink_ultrafast_2015,claassen_dynamical_2017,kitamura_probing_2017}, superconductors 
\cite{raines_enhancement_2015,hoppner_redistribution_2015, Kitamura2016, sentef_theory_2016, knap_dynamical_2016, sentef_theory_2017, kennes_transient_2017, murakami_nonequilibrium_2017, wang_light-enhanced_2018,sheikhan_dynamically_2019,tindall_heating-induced_2019}, 
strongly correlated materials
 \cite{Tsuji2011, tancogne-dejean_ultrafast_2018, walldorf_antiferromagnetic_2019,kennes_floquet_2018,peronaci_enhancement_2020,Takasan2017} and magnetic topological phase transitions \cite{topp_all-optical_2018}. 

A major limitation to Floquet engineering is heating. The generic steady state of an isolated periodically driven many-body system is an infinite-temperature state \cite{dAlessio2014,Lazarides2014}, and although interesting Floquet phases may emerge as prethermal states \cite{Weidinger2017,Canovi2016,Abanin2015,dasari_transient_2018,Herrmann2017,Bukov2015b}, many of the above-mentioned theoretical predictions have not been implemented in solids. Even in cold atom systems, heating can be substantial \cite{Sandholzer2019, Reitter2017, Weinberg2015}. In particular, the qualitatively most interesting effect on many-body interactions is typically achieved by driving a system close to a resonance, like phonon frequency for superconducting pairing \cite{knap_dynamical_2016}, or the Mott gap for the magnetic super-exchange \cite{mentink_ultrafast_2015}, but this near-resonant regime is also where heating is most substantial \cite{murakami_nonequilibrium_2017}.

On the other hand, quantum fluctuations of photon fields in cavities open the possibility to change the properties of matter 
through light-matter coupling without the need of strong lasers.
In particular cavities have the advantage that strong light-matter coupling is in principle achievable, much stronger than the bare coupling in free space. Cavity quantum-electrodynamical environments therefore provide a new paradigm for using light-matter interactions for the creation of effective Hamiltonians with tunable interactions, with intriguing proposals ranging from light-induced superconducting pairing to magnetic super-exchange or ferroelectricity \cite{hagenmuller_cavity-enhanced_2017, sentef_cavity_2018, schlawin_cavity-mediated_2019, curtis_cavity_2019, mazza_superradiant_2019, Andolina2019,kiffner_manipulating_2019, rokaj_quantum_2019, wang_cavity_2019}. 

While such a control of many-body interactions has been  discussed mostly in the deep quantum limit, where vacuum fluctuations alone affect the solid, one can  anticipate that driving the cavity state out of equilibrium implies a continuous crossover to the classical limit of Floquet engineering.
{\color{black}Simply speaking, this crossover is expected to exist because both in the quantum limit and in the Floquet limit the many-body system can exchange photons with the light field, giving rise to induced interactions  between the low energy degrees of freedom. In the quantum limit, an example for effective electronic interactions mediated by a vacuum of bosons can be found in the well-known Bardeen-Cooper-Schrieffer mechanism for phonon-mediated electron-electron attraction, which comes about through the exchange of virtual phonons between electrons. Similarly, induced interaction still emerge when the bosonic field is excited, and populated with states of few photons or phonons or superpositions thereof. In this case both virtual photon emission and absorption contribute to the induced interaction.  Classical laser fields finally correspond to coherent states of high photon numbers, so one can anticipate that the induced interactions in the Floquet Hamiltonian  arise when virtual photon absorption and emission become entirely symmetric. This will be explicitly shown below.}

The basic question to be asked here is whether one can, at strong coupling, achieve with only few photons a similar renormalization of the Hamiltonian as in the classically driven case. Because heating is intrinsically related to the presence of an infinite energy density in the photon system, it should be less relevant if only few photons partake. Naively one might assume that the classically driven Floquet limit is recovered \jl{only} when the cavity is put in a coherent state, but, as we \jl{will} explain in this work, this is not generally true: At strong light-matter coupling, a Hamiltonian similar to the Floquet Hamiltonian can be engineered by putting the cavity in a given photon-number state (with zero expectation value of the driving field), while a coherent state will lead to a more complicated dynamics which is not described by a single effective matter Hamiltonian.

\jl{In this paper, we address this fundamental question by demonstrating the crossover from cavity-coupling to coherent Floquet engineering }
for two important classes of Floquet problems:  (i) The renormalization of tunnelling (dynamical localizaton) \cite{Shirley1965,Dunlap1986}, which underlies the Floquet band-structure control, and (ii), effective induced interactions such as kinetic spin exchange emerging from mobile electrons with Coulomb repulsion, which can be obtained from the Schrieffer-Wolff transformation. The Schrieffer-Wolff transformation is a perturbative framework to derive effective interactions in a sub-Hilbert space, when the rest of the states is projected out; its application includes the derivation of spin models, the $t-J$-model, phonon-mediated electron-electron interactions, and more \cite{bravyi_schriefferwolff_2011}. The Floquet-Schrieffer-Wolff transformation \cite{Bukov2016} is therefore an equally powerful approach to understand the design of induced interactions under periodic driving. Here we present a formulation of the Schrieffer-Wolff transformation of the light-matter Hamiltonian which is in close analogy to the Floquet-Schrieffer-Wolff transformation, and thus shows how the Floquet induced interactions are approached by the induced interactions in the cavity when the photon number is increased 
.

The Schrieffer-Wolff transformation is mathematically similar for different systems, and we investigate it for the paradigmatic example of the spin exchange interaction. The classically driven system has been examined in photo-excited solid-state materials as well as shaken cold-atom systems \cite{mikhaylovskiy_ultrafast_2015, desbuquois_controlling_2017,gorg_enhancement_2018}, and provides an interesting route both for designing exotic spin models \cite{claassen_dynamical_2017} and for the ultrafast control of magnetism \cite{Mentink2017review,Kirilyuk2010}.
 While in the Floquet limit it is possible to reverse the sign of the interaction \cite{mentink_ultrafast_2015} (as experimentally observed in \cite{gorg_enhancement_2018}), vacuum fluctuations alone only reduce the exchange \cite{kiffner_manipulating_2019}. Here we will see that already a single photon can be enough to allow for the sign-reversal of the interaction and almost quantitatively restore the classical Floquet limit. 

This paper is organized as follows. In Sec.~\ref{Sec:floquet-crossover-hop}, we discuss the cavity-coupled Hubbard model and the crossover of the dynamical localization phenomenon into the quantum regime. In Sec.~\ref{Sec:SW}, the Schrieffer-Wolff transformation is discussed for a two-site Hubbard model and a corresponding spin-photon Heisenberg model is derived at large $U$. In Sec.~\ref{Sec:floquet-crossover-spin}, we discuss the high-frequency limit of the cavity-Heisenberg dimer, and consider its crossover from the Floquet driving limit, where the photon number $n$ approaches infinity, to the extreme quantum light regime where only a few photons are present in the cavity. 
Section \ref{Sec:numerics} supplements the previous discussion with a numerical solution of a minimal model for a driven cavity, where cavity photons are created through an external classical driving, and  Sec.~\ref{Sec:conclusion} includes the conclusion and outlook.

\section{Cavity-induced dynamical localization}
\label{Sec:floquet-crossover-hop}

In a certain sense, the link between classically driven Floquet systems and quantum systems is rather straightforward. Consider a Hamiltonian $H^{cl}=\mathcal{H}[Q(t)]$ depending on a classical driving field $Q(t)=A\cos(\Omega t)$. Floquet states are given by a Bloch wavefunction in time, $\psi(t)=u(t)e^{-i\epsilon t}$, where the $u(t)=u(t+T)$  is periodic in time  ($T=2\pi/\Omega$). If expanded in a Fourier series, $u(t)=\sum_n e^{-i\Omega nt } u_n$, the coefficients $u_n$  can be viewed as a wavefunction in a product space $|\alpha,n\rangle $ of matter states $|\alpha\rangle$ and a Floquet index $n$, and the Floquet states are obtained from a solution of the time-independent Schr\"odinger equation with a Hamiltonian $H^{F}_{\alpha,n;\alpha',m}=\frac{1}{T}\int_0^T dt e^{i\Omega (m-n)t} H_{\alpha,\alpha'}(t)$, known as the Floquet matrix. On the other hand, if the drive is replaced by the displacement of a quantum oscillator, $H^{qu}=\mathcal{H}[g(a^\dagger+a)] + \Omega a^\dagger a$, one can project the Hamiltonian on photon numbers. 
\jl{In the following, we will demonstrate} 
that the Floquet Hamiltonian emerges as a classical limit, $H^{qu}(g)_{\alpha,n+m;\alpha',n} \to H^{F}_{\alpha,n+m;\alpha',n}$, when the photon number $n$ is large and the coupling $g$ is small,
\begin{align}
\label{mainlimit}
n\to\infty, \,\,\,g\sqrt{n} \text{~fixed}.
\end{align}
\jl{Indeed, this statement already} 
indicates that at strong light-matter coupling, states with few photons may be sufficient to realize effective Hamiltonians similar to the Floquet Hamiltonian. The structural similarity between the Floquet matrix and cavity quantum electrodynamics has been considered in {\it ab initio} calculations \cite{schaefer_ab_initio_2018,huebener2020}. 

To concretely demonstrate the crossover from classical Floquet driving to the quantum light regime, we consider a 1D 
tight-binding model
coupled to a single cavity mode. We assume the long-wavelength limit, i.e., the photon wavelength is much larger than the system size, so that one can make the dipole approximation. The cavity photon is described by a vector potential $\hat{A} = g \left( \hat{a}^{} + \hat{a}^{\dagger} \right)$, where $g$ is a dimensionless light-matter coupling strength determined by the cavity setup and the operator $a,a^\dag$ annihilate/create a photon in the mode. With the electronic annihilation (creation) operators $c_{j,\sigma}^{(\dagger)}$ acting on site $j$ and spin $\sigma=\uparrow,\downarrow$ and the number operator $\hat{n}_{j,\sigma} = \hat{c}_{j,\sigma}^\dagger \hat{c}_{j,\sigma}^{}$,  the minimal gauge-invariant Hamiltonian is given by
\begin{align}
    \hat{H} =& t_{h} \sum_{i\sigma} \left( \hat{c}_{j,\sigma}^\dagger \hat{c}_{j+1,\sigma}^{} \; e^{i\hat{A}} + \textrm{H.c.} \right) + 
    \Omega \hat{a}^{\dagger} \hat{a}^{}.
    \label{hgejde2kjh2ekln}
\end{align}
Here $t_h$ is the electronic hopping matrix element between neighboring atoms, 
and $\Omega$ denotes the bare cavity photon frequency. This minimal gauge-invariant model can also be derived from the microscopic description under certain circumstance \cite{Li2020} and is in line with the Peierls substitution in the semi-classical limit. It is worth noting that an expansion in powers of $g$ has been used to yield a bilinear term coupling the photonic displacement to the electronic bond current (``paramagnetic term'') in related works \cite{kiffner_manipulating_2019,schlawin_cavity-mediated_2019}, which should be fine in the weak or intermediate coupling limit. For larger coupling, a next-leading term that is second order in $g$ and couples the squared vector potential $\hat{A}^2$ to the kinetic energy of the electrons (``diamagnetic term'') can play a crucial role \cite{kiffner_erratum:_2019}. Even in the weak light-matter coupling limit, a many-photon state may still feature a large amplitude $\langle \hat{A}^2\rangle$, and the perturbation theory breaks down. In fact, this is nothing but the semiclassical limit discussed in the introduction. As we intend to make a bridge between these extreme regimes, we keep the exponential to all orders in the following. 

From the Hamiltonian \eqref{hgejde2kjh2ekln}, taking a proper semi-classical limit at $n\to\infty$ should recover the phenomenology of a Floquet driven system. In the latter case, the cavity mode should be replaced by a coherent electric field $\hat{A}\to A\cos\Omega t$, leading to the Hamiltonian
\begin{align}
\hat{H}_{ cl}&=t_{h} \sum_{j\sigma} \left( \hat{c}_{j,\sigma}^\dagger \hat{c}_{j+1,\sigma}^{} \; e^{iA\cos\Omega t} + \textrm{H.c.} \right),
\end{align}
where $A,\Omega$ are the amplitude and frequency of the periodic external field. The effective Floquet Hamiltonian in the high-frequency $\Omega\to\infty$ limit is known to have a renormalized hopping \cite{Dunlap1986,Bukov2015}
\begin{align}
t_{h}^{F}(A)=t_{h}J_0(A),
\end{align}
where $J_0(x)$ is the zeroth Bessel function of the first kind. 

\begin{figure}[tbp]
\centering
\includegraphics[width=0.45\textwidth]{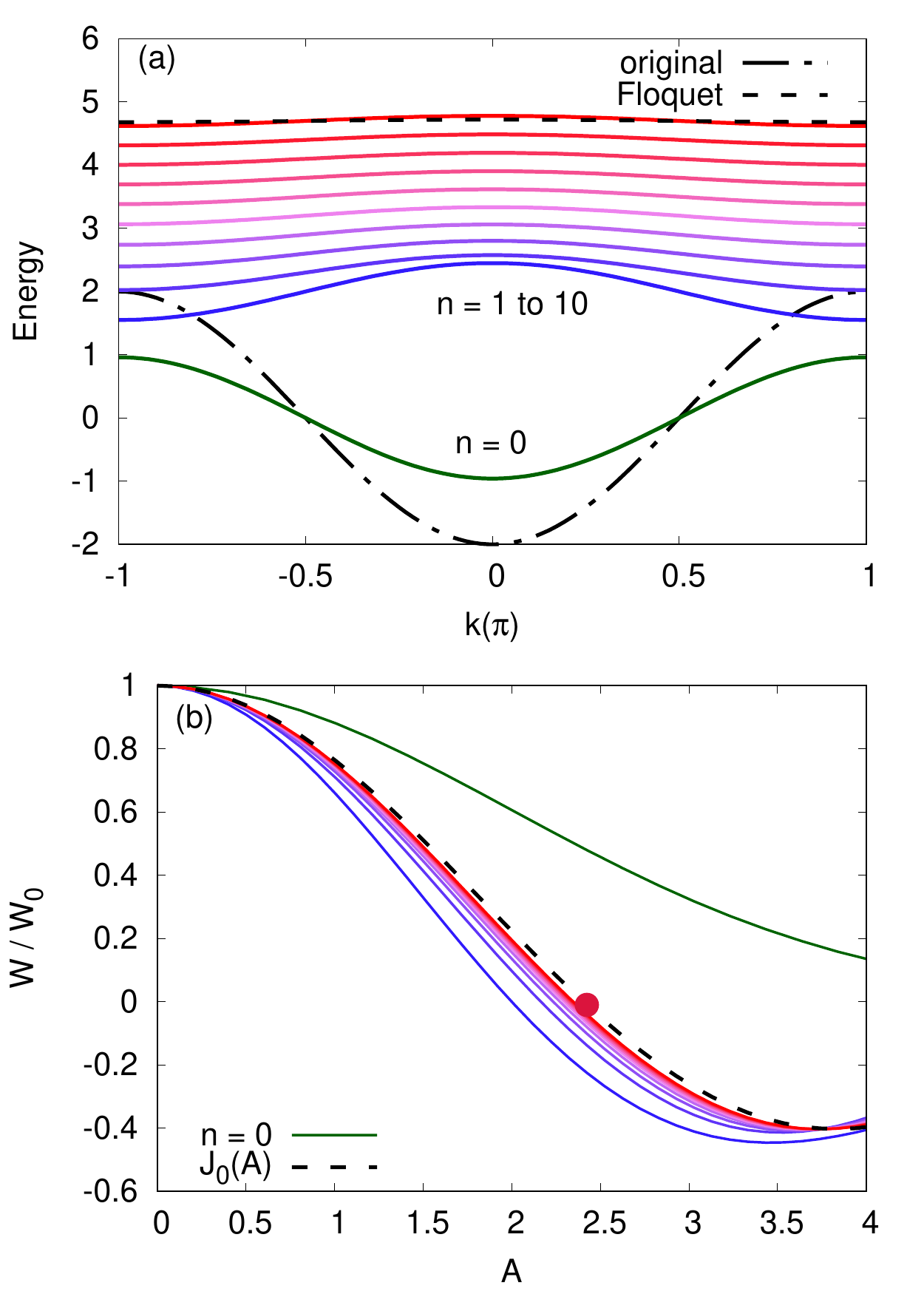}
\caption{{\bf Cavity-Floquet crossover for dynamical localization.} (a) The evolution of non-interacting energy band for increasing photon number $n=0,\ldots,10$, with $2g\sqrt{n}=2.42$ fixed ($2g=2.42$ for $n=0$). The black dot-dashed line indicates the undriven band ($-2t_h\cos k$) and the dashed line is the Floquet case. The curves are shifted vertically for visibility. (b) The relative bandwidth $W/W_0$ with $W_0=2t_h$ as a function of $A=2g\sqrt{n}$ and $n=0,\ldots,10$. The red dot labels the value $A=2.42$ for the curves in panel a), which is the first zero of the Bessel function $J_0$. For $n=0$, $A=2g$ is adopted. In both (a) and (b) the same color scheme is used for indicating $n$.}
\label{fig:dl}
\end{figure}

We now concentrate on the corresponding renormalization of hopping $t_h$ due to the cavity photons in the high frequency limit. Specifically, we perform a unitary transformation $\hat{U}(t)=\exp(i\Omega a^\dag a t)$ on the cavity-lattice Hamiltonian~\eqref{hgejde2kjh2ekln} to enter the rotating frame
{\color{black} (equivalent to going to the interaction picture with respect to $i\Omega a^\dag a$)}. 
This removes the term $\Omega a^\dag a$, and leads to the replacement  $a\to a e^{-i\Omega t}$. The transformed Hamiltonian $\hat{H}_{rot}(t)=\hat{U}(t)[\hat{H}-i\partial_t]\hat{U}^\dag(t)$ is then periodic in time $\hat{H}_{rot}(t+2\pi/\Omega)=\hat{H}_{rot}(t)$, and one can perform a high-frequency expansion \cite{Bukov2015}. The effective Hamiltonian at the lowest order, $\hat{H}_{\rm eff}=\frac{1}{T}\int _0^T dt \hat{H}_{rot}(t)$, is 
\begin{align}
\hat{H}_{\rm eff}&=t_{h} \sum_{\sigma} \left(\hat{J}_{ h}(2g)\hat{c}_{j,\sigma}^\dagger \hat{c}_{j+1,\sigma}^{} + \textrm{H.c.} \right),
\end{align}
where $\hat{J}_{ h}(x)=e^{-x^2/8}\sum_{k=0}^n(ix/2)^{2k}(a^\dag)^ka^k/k!^2$.
(For details, see appendix \ref{sec:appA1}).
When $n$ photons are present in the cavity, one can see that the electronic hopping is renormalized by a factor $J^{(n)}_{ h}(x)=\langle n|\hat{J}_{h}(x)|n\rangle$, i.e., $t_h^{(n)}(g)=t_{h}J^{(n)}_{h}(2g)$. Finally, under the classical limit defined in Eq.~\eqref{mainlimit}, one can show that this renormalized hopping approaches the Floquet limit,
\begin{align}
\label{limitdynlovc}
\lim_{n\to\infty}
t^{(n)}_{h}(g)
=
t_{h}^{F}(A).  
\text{~~~}(2g\sqrt{n}=A\text{~fixed)}.
\end{align}

The Floquet-cavity crossover for dynamical localization is explicitly illustrated in Fig.~\ref{fig:dl}. Panel a) shows the effective energy band with dispersion $-2t^{(n)}_{h}\cos(k)$ for different $n$ ($A=2g/\sqrt{n}$ fixed). In the large $n$ limit the energy band becomes almost flat, which is consistent with the Floquet limit where $t_h^F\propto J_{0}(A=2.42)\approx0$ for the given parameters. Figure~\ref{fig:dl}b shows the renormalized hoppings $t_h^{(n)}$ as a function of the coupling. In the dark cavity ($n=0$), the factor $J^{(0)}_{ h}=e^{-g^2/2}$ always leads to a reduced bandwidth $t_h^{(0)}<t_h$, while the Floquet drive allows for a flipping of the band, with interesting consequences in interacting systems \cite{Tsuji2011}. With more photons in the cavity ($n>0$), both emission and absorption of virtual photons contribute to $J^{(n)}_{ h}$. It is interesting to see that a single photon is already sufficient to flip the band and thus  correct for the qualitative difference between the dark cavity and the classically driven system.

As emphasized  in the introduction, the Floquet Hamiltonian is recovered when the cavity is in a number state, and no coherent state is assumed. At small coupling and large photon number $n$, however, the photon number state and a coherent state would nevertheless give the same result, because the coherent state has a small variance in the photon number $\Delta n \ll \langle n\rangle$. In the classical limit, therefore, the Floquet Hamiltonian can be realized by coherent driving. At large coupling and small photon number, in contrast, the coherent state would give an dynamics which cannot be described by a single tunnelling matrix element at all. This will be illustrated in more detail for the following example.

\section{Schrieffer-Wolff transformation in the cavity}
\label{Sec:SW}

In this section we extend the Hamiltonian \eqref{hgejde2kjh2ekln} by a local Hubbard interaction $U$,
\begin{align}
    \hat{H} =& t_{h} \sum_{j\sigma} \left( \hat{c}_{j,\sigma}^\dagger \hat{c}_{j+1,\sigma}^{} \; e^{i\hat{A}} + \textrm{H.c.} \right) + 
    \nonumber \\ &+
     U \sum_j \hat{n}_{j,\uparrow} \hat{n}_{j,\downarrow} + \Omega \hat{a}^{\dagger} \hat{a}^{}.
    \label{hgejde2kjh2ekln001}
\end{align}
{\color{black}
We will take this as an example to discuss the cavity to Floquet crossover regarding induced interactions in a low-energy space. For this purpose, we focus on the limit $U\gg t_h$ at half filling (one electron per site). In this limit, it is well known that the effective low-energy Hamiltonian without coupling to the electromagnetic field is obtained by projecting out configurations  which contain doubly occupied and empty sites (Schrieffer-Wolff transformation). The resulting low energy space has one spin-1/2 at each lattice site, and the effective  Hamiltonian is a Heisenberg model $H=J_{ex}\sum_{i} {\bm S}_i {\bm S}_{i+1}$ with exchange interaction $J_{ex}=4t_h^2/U$ (${\bm S}_j$ are canonical spin-$\frac12$ operators). In the following we  include the coupling to the electromagnetic field, but still focus on the limit $U\gg t_h$ and ask how the low-energy spin model and the induced spin interactions are  modified by either classical driving (replacing $\hat A \to A\cos(\Omega t)$ as in the previous section) or by coupling to the quantum field, and how the two limits are related.  For this purpose it is sufficient to consider a minimal Hubbard lattice of two sites $j=1,2$ (Hubbard dimer).}

{\color{black}
With the coupling to the cavity mode, the relevant Hilbert space at $U\gg t_h$, after projecting out electronically excited doubly occupied states, will contain both  spins and  photons, and the effective Hamiltonian is therefore a spin-photon model. Below we will derive a suitable Schrieffer-Wolff transformation for the electron-photon Hamiltonian (7) to show that the Hubbard dimer reduces to the spin-photon Heisenberg model}
\begin{align}
\label{vghhssssh}
H = (\bm S_1 \bm S_2 -\tfrac12)\mathcal{J}[a^\dagger,a] + \Omega a^\dagger a,
\end{align}
where 
the exchange interaction becomes an operator $\mathcal{J}[a^\dagger,a]$ acting on the photon states.  
{\color{black}
In deriving this expression, we only assume the absence of (multi-photon) resonances, i.e., $n\Omega\neq U$ for any integer $n$. }
It is convenient to 
separate $\mathcal{J}[a^\dagger,a]$ into photon number transitions,
\begin{align}
\mathcal{J}[a^\dagger,a]
&=
\mathcal{J}_0[a^\dagger,a]
+\sum_{m=1}^\infty
\Big(
(a^\dagger)^{2m}\mathcal{J}_{2m}[a^\dagger,a] + 
h.c.\Big),
\label{gggegerr2}
\\
\mathcal{J}_{2m}
&=
J_{\rm ex}
\sum_{c=0}^\infty
g^{2c+2m}(a^\dagger)^{c}a^c\,
\mathcal{L}_{c,m}(g,\bar \omega),
\label{debwelwe}
\end{align}
where $\mathcal{J}_n$ are (normal ordered) hermitian operators which are diagonal in the photon number,  $\bar\omega=\Omega/U$, and the overall scale is the bare kinetic exchange interaction $J_{\rm ex}=\frac{4t_{h}^2}{U}$. 
 Note that only even photon number transitions have non-vanishing matrix elements.  
The function $\mathcal{L}_{c,m}(g,\bar \omega)$ contains the dependence of the exchange interaction on frequency.
Its precise form is given in the appendix in Eq.~(A59). It is a  smooth function of $\Omega$ and $g$, apart from divergencies at the resonances $U=n\Omega$, with integer $n$.
 
\begin{figure}[tbp]
\centering
\includegraphics[width=0.45\textwidth]{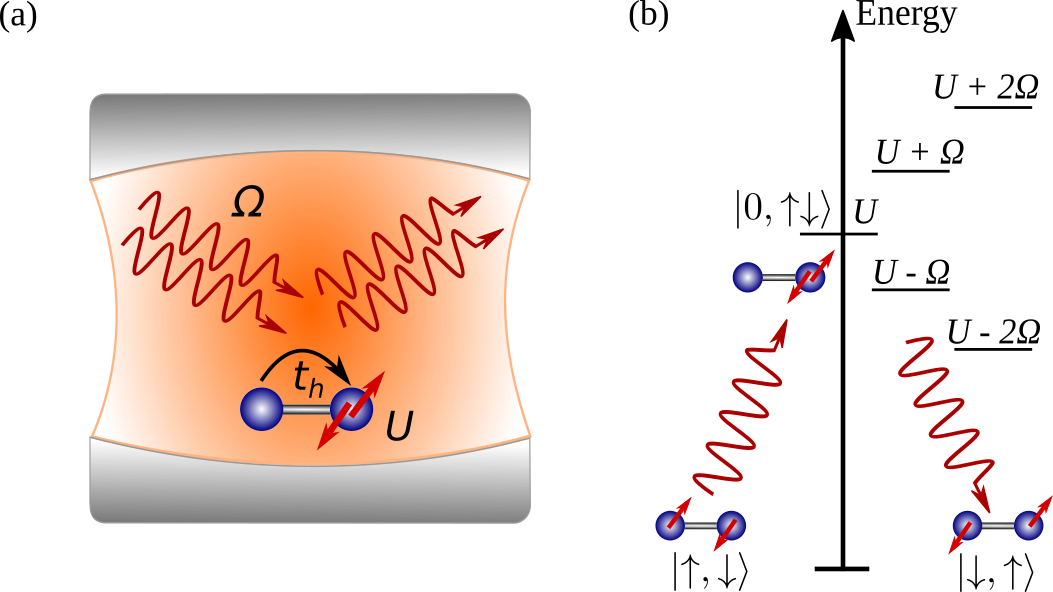}
\caption{{\bf The Hubbard dimer coupled to a photon mode.} 
(a) The sketch of a Hubbard dimer coupled to the cavity photon mode. (b) The energy structure of a cavity coupled Hubbard dimer at the strong coupling $U\gg t_h$ limit. The two lattice sites exchange spins through virtual processes visiting different photon number sectors. 
}
\label{fig:setup}
\end{figure}

 The above equations constitute a central result of this paper. The spin-photon Hamiltonian has a similar form to the Floquet-engineered spin Hamiltonian described below, but describes the full dynamics of the spin and photon-coupled system. In Eq.~\eqref{debwelwe} the non-resonance condition is assumed $U\ne n\Omega$, but no assumption is made for the relation between $\Omega$ and the low energy scale $J_{\rm ex}$, leaving the full photon dynamics intact. Thus, the Hamiltonian describes both the photon-engineered spin dynamics, and the modification of photon states due to the presence of magnetic degrees of freedom. In the following we briefly  discuss the renormalization of the  photon states, and then turn to the cavity-engineering of the spin-exchange coupling.

{\color{black}
Below we contrast these  results to the Floquet-driven system.  The Floquet spin Hamiltonian at $U\gg t_h$ has been obtained in the same spirit as the Heisenberg model above, by assuming that the non-resonant driving does not generate charge excitations, and one can thus project out doubly occupied states from the Floquet Hamiltonian  \cite{mentink_ultrafast_2015}. Alternative derivations have been formulated in various  ways, including operator-based Floquet Schrieffer-Wolff transformations \cite{Bukov2016}, time-dependent Schrieffer Wolff transformations  \cite{eckstein_designing_2017}, or a resummation of the high-frequency expansion \cite{itin2015}.
}
For $U\gg t_h$, the low-energy physics of the Floquet-driven system is described by the Floquet Heisenberg model $\hat{H}^{F}=J^{F}_{\rm ex}{\bm S}_1{\bm S}_2$, with exchange interaction given by \cite{mentink_ultrafast_2015}
\begin{align}
J^{F}_{\rm ex}=J_{\rm ex}\sum_{\ell}\frac{J_{|\ell|}(A)^2}{1-\ell \Omega/U},
\label{Jfl}
\end{align}
with $J_{\rm ex}=4t_h^2/U$ being the exchange interaction of the undriven system and $J_{\ell}(A)$ being the $\ell$th Bessel function. This expression can be schematically explained with the multiple Floquet sectors with energy $U+\ell\Omega$, for $\ell\in\mathbb{Z}$, contributing to the virtual spin exchange process, as shown in Fig.~\ref{fig:setup}(b). Note that $\Omega\gg J_{\rm ex}$ is needed to exclude real photon emission/absorption in the effective model \eqref{Jfl}. In the undriven case $A=0$, the usual Heisenberg model is restored as $J_{\ell}(0)=\delta_{0\ell}$. 

Analogously, we consider the limit $U\gg t_h$ of the cavity-Hubbard dimer \eqref{hgejde2kjh2ekln001}. In this limit, the induced magnetic interaction ($\bm{S}_1\bm{S_2}$)  emerges due to electron hopping between neighboring lattice sites with an intermediate excited state. 
For example, one electron (say spin up) at site 1 can hop to its neighbor 2 and form a spin singlet (the doublon) at 2. If $|\ell|$ photons are absorbed (emitted) in this process, the intermediate state then has excess energy $U\pm|\ell|\Omega$, respectively. The other electron (spin down) can eventually hop back to site 1, with the net result of exchanging the two spins. If the high-frequency limit is taken, the absorbed (emitted) photon have to be emitted (absorbed) back, and the system must go back to the original photon-number state. This is in parallel with the Floquet scenario as discussed above. However, in that case the energy $\ell\Omega$ is borrowed from (or lent to for negative $\ell$) the classical driving field, instead of the quantized cavity levels.

We briefly describe here how to systematically derive the effective model Eq.~\ref{vghhssssh} and refer to Appendix ~\ref{Sec:dervation} for details. One again applies the unitary transformation $\hat{U}=\exp(i\Omega a^\dag a t)$ and separate the Hilbert space into sectors $\mathcal{H}_0$ and $\mathcal{H}_1$ without and with charge excitations (doublons, holes), respectively. No assumption is made on $\Omega$, and both sectors may contain an arbitrary number of photons. It is then possible to perform a subsequent time-periodic unitary transformation, analogous to the Floquet Schrieffer-Wolff transformation, such that the coupling matrix elements between the two sectors are small in $t_{h}/U$, and after that project to the sector $0$. This procedure is essentially a generalized Schrieffer-Wolff transformation in the electron-photon Hilbert space. Details of this cavity Schrieffer-Wolff transformation can be found in Appendix ~\ref{Sec:dervation}.
 
  \begin{figure}[tb]
\centering
	 \includegraphics[width=0.45\textwidth]{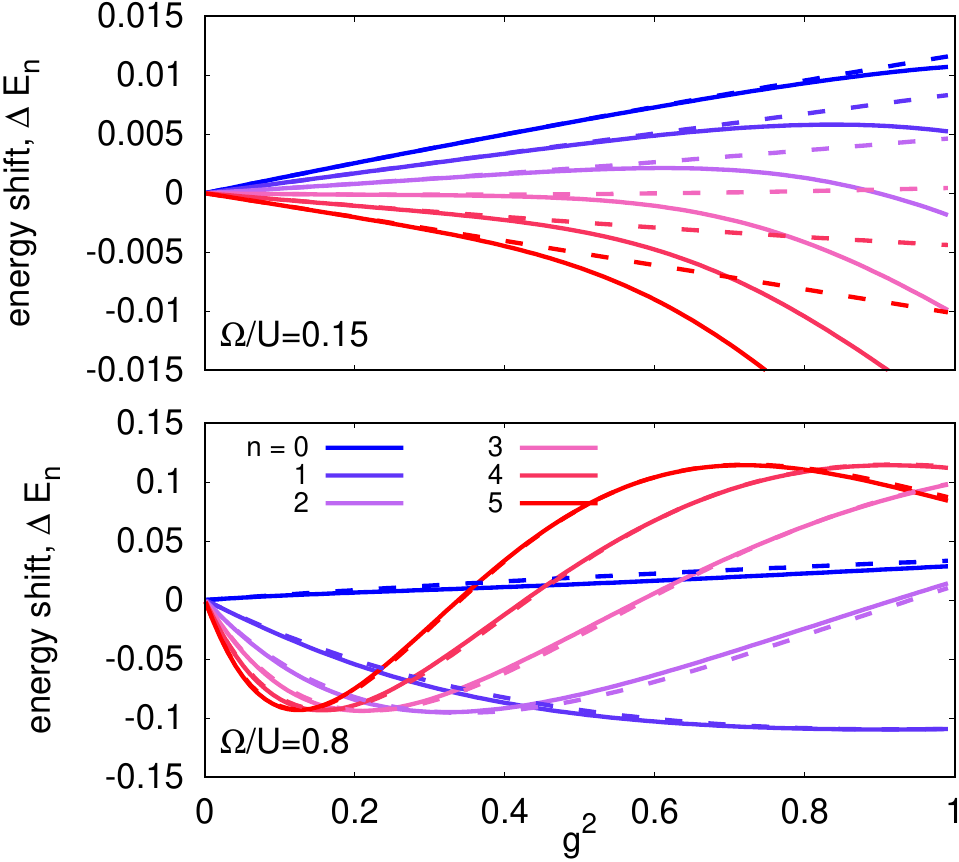}
\caption{
	{\bf Energy shift $\Delta E_n = E_n - n\Omega$ due to the light-matter coupling for the singlet $S_{tot}=0$.}  The dashed lines correspond to an evaluation in the high-frequency limit.} \label{fig:evalsinglet}
\end{figure}

 \paragraph*{Squeezed photon states}
 In general, the photon states are coupled to magnetic excitations in the spin-photon Heisenberg model. For the dimer, the eigenfunctions are obtained in the form $|S_{tot},S_z\rangle|\Psi_m\rangle$, where $\left|S_{tot},S_z\right\rangle$ is the spin wave function, which is a singlet ($S_{tot}=0$) or triplet ($S_{tot}=1$), and $\left|\Psi_m\right\rangle$ is the eigenfunction of the operator $\Omega a^\dagger a - (1-S_{tot})\mathcal{J}[a^\dagger,a]$. The operator $\mathcal{J}[a^\dagger,a]$ contains various photon nonlinearities. For example, it can be readily evaluated at weak coupling $g\ll 1$
\begin{align}
\frac{\mathcal{J}_{0}}{J_{\rm ex}}
&=1
-g^2\frac{\bar\omega}{1+\bar\omega}
+
g^2
a^\dagger a
\frac{2\bar\omega^2}{1-\bar\omega^2}+ \mathcal{O}(g^4),
\\
\frac{\mathcal{J}_{2}}{J_{\rm ex}}
&=
g^{2}
\frac{\bar\omega^2+2\bar\omega^4}{(1-4\bar\omega^2)(1-\bar\omega^2)}+ \mathcal{O}(g^4),
\label{phpononhgefak}
 \end{align}
so that the cavity wave functions for $S_{tot}=0$ are squeezed, and the cavity frequency is shifted. Figure \ref{fig:evalsinglet} shows the energy shift relative to the free photon $\Delta E_n = E_n - n\Omega$. The results are obtained by diagonalizing  $\Omega a^\dagger a - (1-S_{tot})\mathcal{J}[a^\dagger,a]$. At small $g$ the shift of the cavity frequency is proportional to $g^2$. For larger $g$ nonlinearities (photon self-interaction) set in. The results obtained in the high-frequency limit, omitting the photon-non-diagonal terms $\sim \mathcal{J}_{n\neq 0}$ in the Hamiltonian, are shown in the figure with dashed lines.
 The high-frequency expansion is reasonable when $\Omega \gg J_{\rm ex}$ (lower panel). In contrast, when $\Omega\sim J_{\rm ex}$, the photon ground wave function becomes a squeezed state, with admixtures from $|n=2,4,...\rangle$. 

It is worth noting that the Schrieffer-Wolff transformation can be performed for the Hubbard model on an arbitrary lattice,
\begin{align}
\label{vghhsssshfull}
H = \sum_{\langle rs \rangle}(\bm S_r \bm S_s -\tfrac12)\mathcal{J}_{\langle rs \rangle}[a^\dagger,a] + \Omega a^\dagger a,
\end{align}
where $\mathcal{J}_{\langle rs \rangle}$ is determined by the exchange operator of the two-site model with the hopping $t_{h}$ and coupling $g$ for the given bond. On an arbitrary lattice, 
spin excitations can be created through photon absorption, or vice versa.
For example, if the polarization is such that only bonds along one direction of the lattice are affected, the Hamiltonian gives rise to two-magnon two-photon scattering terms. The discussion of this rich physics is left for future works. 

\section{The Cavity-Floquet crossover}
\label{Sec:floquet-crossover-spin}
\subsection{Floquet crossover of the photon-number states}

We now turn to the cavity-engineering of the spin dynamics in the high-frequency limit $\Omega \gg J_{\rm ex}$ where one can project out the creation and annihilation of real photon excitations. We emphasize that, distinct from the case of dynamical localization, the frequency does not have to be high compared to the energy scales $U$ or $t_{h}$ which have already been removed from the Hamiltonian. In the high-frequency limit one could perform another unitary transformation, which rotates away the transition matrix elements $\sim \mathcal{J}_{n\neq0}$ between different photon number sectors. This would result in corrections of order $J_{\rm ex}^2/\Omega \ll J_{\rm ex} $ to the photon-diagonal exchange Hamiltonian, which are omitted \footnote{For example, if $U$ and $\Omega$ are of the same order, $J_{\rm ex}^2/\Omega$ is of order $t_{h}^4/U^3$, and should thus be omitted consistent with the lowest order expansion in $t_{h}/U$.}. The resulting exchange operator thus becomes photon diagonal, and the full Hamiltonian becomes
\begin{align}
H
&=
\sum_{n}
|n\rangle\langle n|
\big(H^{\rm spin}_n+n\Omega\big),
\end{align}
with $
H^{\rm spin}_n=
(\bm S_1 \bm S_2 -\tfrac12)
J_{\rm ex}^{(n)}$, and an exchange interaction $J_{\rm ex}^{(n)}=\langle n|\mathcal{J}_{0}|n \rangle$.
%
%
Analogous to the dynamical localization [Eq.~\eqref{limitdynlovc}], one can now show explicitly that in the classical limit \eqref{mainlimit}, the exchange interaction $J_{\rm ex}^{(n)}$ approaches the Floquet result \eqref{Jfl} (see appendix A.1 for details),
\begin{align}
\label{nsjsj}
 J_{\rm ex}^{(n)}(g) \to 
 J_{\rm ex}^{F}(A) 
\,\,\,\,\, (n\to\infty, A=2\sqrt{n}g \text{~fixed}).
\end{align}

\begin{figure}[tbp]
\centering
\includegraphics[width=0.45\textwidth]{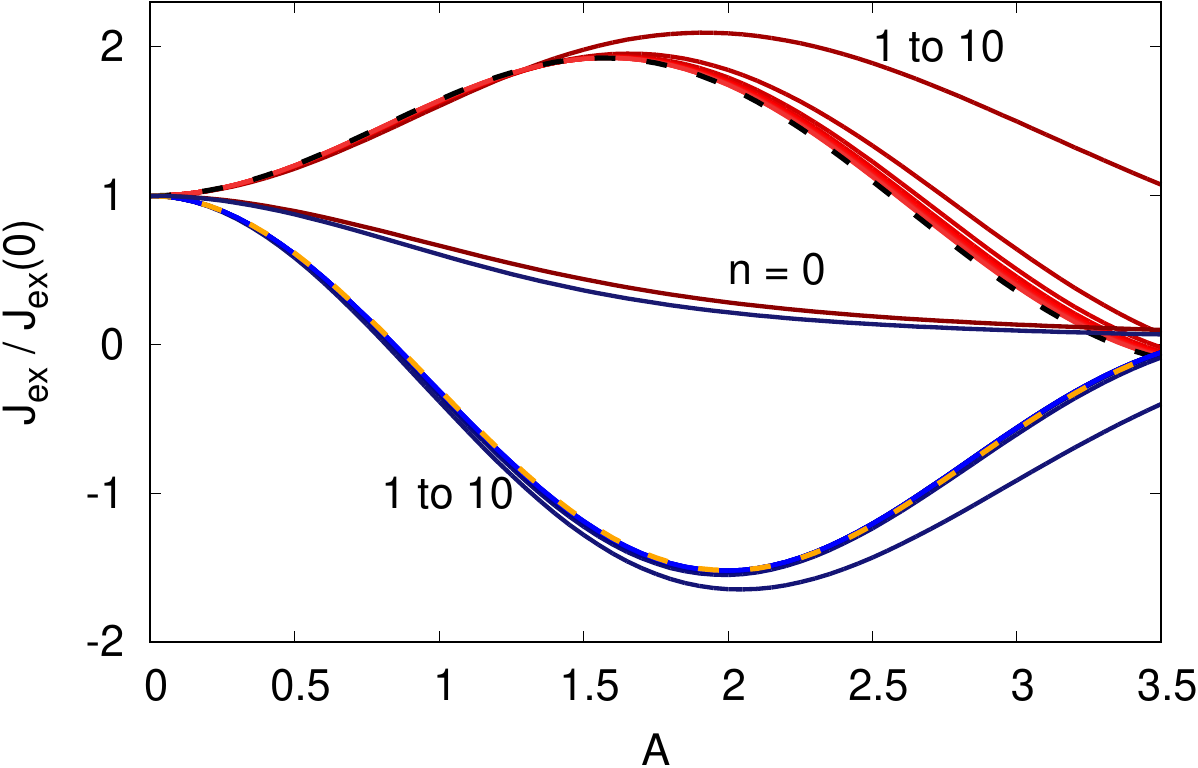}
	\caption{{\bf
Exchange interaction $J_{\rm ex}^{(n)}$  
	in the $n$-photon state as a function of $A$ for $n=1,...,10$.} The vertical axis is in units of $J_{\rm ex}(0)$ at coupling $g=0$. The curves with colors from dark to light red correspond to $\Omega=0.8U$ (red-detuned) and those with dark to light blue correspond to $\Omega=1.2U$ (blue-detuned). The lightness of the color indicates the photon number $n$ with darkest ones denote $n=0$. The dashed black (orange) line shows the Floquet result \eqref{Jfl} for $\Omega=0.8U$ ($\Omega=1.2U$). For the dark cavity ($n=0$) case, $J_{\rm ex}^{(0)}$ is plotted as a function of the coupling $g=A$. }
	\label{fig:wjw}
\end{figure}

The behavior of $J^{(n)}_{\rm ex}$, which quickly converges to the Floquet limiting curves (dashed lines) as $n$ rises, is systematically demonstrated in Fig.~\ref{fig:wjw}. For small $A$, the red-detuned ($\Omega<U$) cavity results in an enhanced $J_{\rm ex}$ while the blue-detuned ($\Omega>U$) cavity leads to a reduced $J_{\rm ex}$, and eventually a sign change. In the case of a completely dark cavity ($n=0$), $J^{(0)}_{\rm ex}$ is suppressed for both red and blue-detuned frequencies, and is thus qualitatively different from the Floquet limit $J^{fl}_{\rm ex}$. 
Similar to the case of dynamical localization, however, a single photon is already sufficient to resolve these qualitative differences, and in particular restore the possibility to flip the sign of the exchange interaction. Also quantitatively, we observe a rather fast convergence of $J_{\rm ex}^{(n)}$ with the photon number.

Note that the qualitative enhancement and suppression of $J_{\rm ex}$ in the red- and blue-detuned cases are consistent with the result of Ref.~\onlinecite{kiffner_erratum:_2019} based on an expansion of the light-matter coupling truncated at the quadratic ``diamagnetic" term. The suppression in the blue-detuned case is, in particular, missing in the linear coupling approximation (``paramagnetic" term) \cite{kiffner_manipulating_2019}. 


\subsection{The spin dynamics in the high-frequency limit}


While at strong coupling the Floquet exchange is recovered with only few photons, it should be emphasized that putting the cavity in a  coherent state does not necessarily recover the Floquet result. Instead, the resulting dynamics of the spin subsystem cannot be described by a single spin Hamiltonian at all. In this section we illustrate this fact with a simple precessional spin dynamics:
 We prepare the system in a state 
\begin{align}
\label{heeeless}
|\Psi\rangle = \left|\uparrow,\downarrow\right\rangle \sum_{n}b_n|n\rangle,
\end{align}
where the spins form a Neel state $\left|\uparrow,\downarrow\right\rangle$, which is a singlet-triplet superposition, and the cavity is in an arbitrary state. Then we find for the precessional motion of the spin on site $1$,
\begin{align}
\label{ggegek;;e}
\langle S_1^z(t)\rangle = \frac{1}{2}\sum_{n=0}^\infty |b_n|^2 \cos(J_{\rm ex}^{(n)}t).
\end{align}

The behavior \eqref{ggegek;;e} of the cavity-driven Heisenberg dimer is therefore apparently different from the classical Floquet engineering. In the latter case, upon projecting out high-frequency processes, the Neel state  would precess with a single renormalized frequency $J^{F}_{\rm ex}$,
\begin{align}
\label{ggegek;;e01}
S_1^{z,F}(t) = \frac{1}{2} \cos(J_{\rm ex}^{F}t).
\end{align}
This discrepancy is somehow expected as the coupling to a classical field $A\cos\Omega t$ is fundamentally distinct from the coupling to a few discrete quantum states, where multiple precessional frequencies can emerge out of the coupling to a plethora of discrete levels.

\begin{figure}[htp]
\centering
	\includegraphics[width=0.45\textwidth]{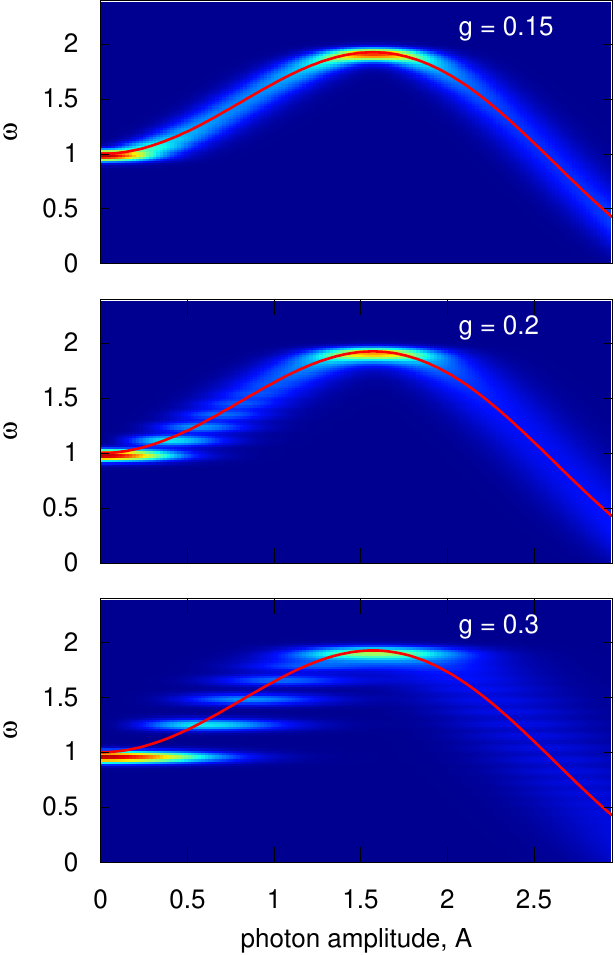}
\caption{{\bf
Visualization of the crossover: Fourier analysis of the spin dynamics.} The Fourier spectrum of the spin-precession modes is shown for a coherent state with $\bar{N}=(A/2g)^2$, as a function of the amplitudes $A$. The color is scaled in units of $J_{\rm ex}=4t_h^2/U$. The red line is $J_{\rm ex}^{F}(A)$. For large $g$, only few photons are needed to get the same modification of $J_{\rm ex}$, and the discreteness of the photon states becomes apparent. The frequency is $\Omega=0.8U$.
}
\label{fig:wjw02}
\end{figure}

To make a clear connection with the Floquet-driven case, we suppose the cavity is prepared in a coherent state with a mean number $\bar N$ of photons, i.e. 
\begin{align}
b_n^2=e^{-\bar N} \frac{\bar N^n}{n!}.
\end{align}
Note that, by identifying $A\cos(\Omega t)$ with $\left\langle\bar{N}\right|\hat{A}\left|\bar{N}\right\rangle(t)$ in the free field limit, one obtains the semiclassical correspondence $A\equiv 2g\sqrt{\bar{N}}$. In Fig.~\ref{fig:wjw02}, we examine the Fourier spectrum of the spin dynamics \eqref{ggegek;;e} for a coherent state with $\bar{N}=(A/2g)^2$: $
S(\omega,A)=\sum_n b_n^2\phi(\omega-J^{(n)}_{\rm ex})$, where $\phi(x)=e^{-x^2/\Delta^2}$ is a broadened $\delta$ peak with frequency resolution $\Delta=0.05$. At different cavity coupling $g$, we plot the Fourier spectrum  and compare it to the Floquet exchange $J^F_{\rm ex}$. The large $\bar{N}$ limit is obtained by taking a small $g$ with a fixed amplitude $A$. It is evident that, with a relatively large $\bar{N}$ (such as $g=0.15$ in the figure), the Fourier spectrum fits very well with the Floquet limit. This is expected as in the semiclassical limit, the cavity-induced modification of $J_{\rm ex}$ should be consistent with the Floquet theory. For larger $g$, at given $A=2g\sqrt{\bar{N}}$, the photon number is small and we observe quantized frequency plateaus in the Fourier spectrum. The plateaus are generally in the vicinity of the Floquet curve $J^{F}_{\rm ex}(A)$. It is especially intriguing to see that, in the strong light-matter coupling limit, even with few photons, ($\bar{N}\lesssim6$ at $A\lesssim 1.5$ for $g=0.3$) the discrete frequency levels still follow closely the Floquet curve $J^{F}_{\rm ex}(A)$. This extends the Floquet-like engineering into the few-photon, or extreme quantum light regime. As $A$ increases, the photon number increases for fixed $g$, and the plateaus become denser and eventually merge into a continuum in the large $\bar{N}$ limit. 
For a coherent state $\left|\bar{N}\right\rangle$, the standard deviation of $n$ is $\sqrt{\bar N}$ and thus the sum \eqref{ggegek;;e} is dominated by the term $\frac{1}{2} \cos(J_{\rm ex}^{(\bar N)}t)$. Therefore, the conventional Floquet engineering is restored in the coherent driving limit. We conclude that the two seemingly distinct situations, coupling to coherent driving and to a photon-number state, both converge to the Floquet-driven scenario in the semi-classical limit. The above discussion also shows that the modification of exchange interaction can be generalized to an arbitrary photon occupation provided $|b_n|^2$ is sharply centered around $n\sim\bar{N}$, without the assumption of a coherent or even a pure cavity state.

\section{Driving the cavity}
\label{Sec:numerics}
Using the generalized Schrieffer-Wolff transformation, we have shown that both coherent states and photon-number states result in Floquet-like modifications of the magnetic dynamics. In more realistic experiments, the cavity would be driven to an excited state by some external field. In addition, the cavity is generically open and dissipative. In the following we consider a minimal setup of a driven cavity, where the cavity is originally prepared in the ground state (a dark cavity) and then is acted on by a time-dependent external laser field $f(t)$ linearly coupled to the photon mode
\begin{align}
    \hat{F}(t) &= f(t) \left( \hat{a}^{} + \hat{a}^{\dagger} \right),
\end{align}
so that the time evolution is determined by the total Hamiltonian
\begin{align}
    \hat{H}(t) &= \hat{H} + \hat{F}(t).
\end{align}
A closed and isolated cavity is still assumed, so that no dissipation or photon leakage is present in the time-evolution. Below we use a driving field
\begin{align}
    f(t) &= F \sin(\Omega_{\text{dr}}t),\quad (t>0)
\end{align}
with driving frequency $\Omega_{\text{dr}}$ and amplitude $F$. 

We perform an exact time evolution with a truncated bosonic Hilbert space \cite{sentef_light-enhanced_2017}
for the the case of a Hubbard dimer. In order to specifically address the photodressing effects on the effective magnetic exchange interaction $J_{\text{eff}}$ we compute the local double-time spin-spin correlation function on the first site equal to the one on the second site,
\begin{align}
    \chi(t,t') &= \langle \psi(t) | S_1^z \hat{U}(t,t') S_1^z | \psi(t') \rangle,
\end{align}
where 
\begin{align}
    S_1^z &= \frac12 \left( \hat{n}_{1,\uparrow} - \hat{n}_{1,\downarrow} \right),
\end{align}
$|\psi(t)\rangle$ denotes the wave function at time $t$, and $\hat{U}(t,t')$ $=$ $\mathcal{T} \exp(-i \int_{t'}^t \hat{H}(s) ds)$ is the unitary time evolution operator with time ordering $\mathcal{T}$ that propagates the wave function from the initial ground state $|\psi(t=0)\rangle$ $=$ $|\psi_0\rangle$ of the undriven Hamiltonian to the time-evolved state,
\begin{align}
    |\psi(t)\rangle &= \hat{U}(t,0) |\psi_0\rangle.
\end{align}
From the double-time response function we compute a time- and frequency-resolved spin susceptibility 
\begin{align}
    \chi(\omega,t_0) &= \int dt \int dt' s(t,t_0) s(t',t_0) \chi(t,t'),
    \label{eq:susc}
\end{align}
with probe envelope function
\begin{align}
    s(t,t_0) &= \exp\left(-\frac12 (t-t_0)^2/\sigma^2\right)
\end{align}
with probe duration $\sigma$ and probe time $t_0$. 
Note that for a pure Heisenberg dimer, $\chi(t,t') \propto \cos(J_{\rm ex}t)$, and therefore 
$\chi(\omega)\propto \exp\left(-\frac12 (\omega -J_{\rm ex})^2 \sigma^2\right)$ has a single peak at $\omega>0$ which directly measures $J_{ex}$.

Below we show results for runs up to $t_\text{max} = 200$, probe time $t_0 = 100$ and probe duration $\sigma = 36$. The units are chosen such that the hopping $t_{h} = 1$ sets the unit of energy, and correspondingly times are measured in units of $\hbar/t_{h}$. We note that $\hbar = 0.658 eV \times \text{fs}$, implying that for $t_{h} = 1 \text{eV}$ the time unit is $0.658 \text{fs}$. Throughout we fix $U/t_{h} = 8$ to be in a relatively strong-coupling limit, for which the spin exchange interaction is perturbatively given by $J_{\text{ex}}$ $=$ $\frac{4 t_{h}^2}{U}$ $=$ $0.5$. We employ driving frequencies $\Omega_{\text{dr}}$ on resonance with the bare cavity mode frequency $\Omega$ and choose these frequencies to be well above the $J_{\text{ex}}$ scale, but on the order of $U$. 
\begin{figure}[htp]
\centering
\includegraphics[width=0.5\textwidth]{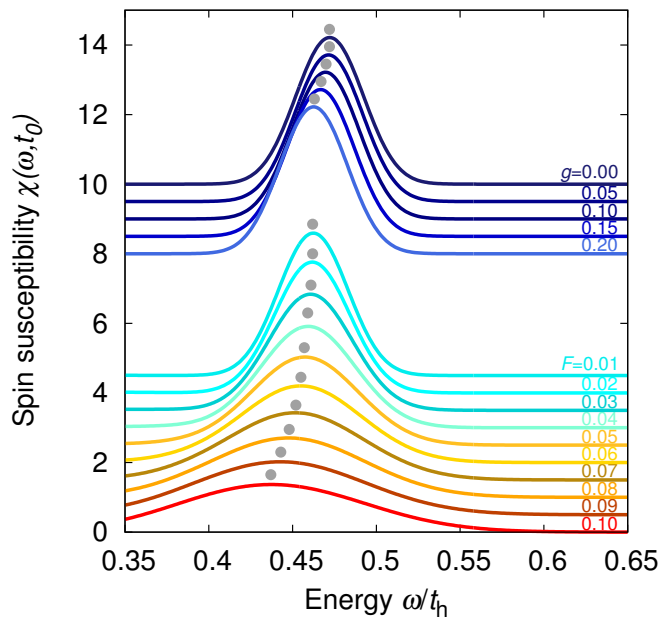}
\caption{{\bf Light-matter tuning of effective exchange interaction at high frequency.} Evolution of frequency-dependent spin susceptibility (Eq.~(\ref{eq:susc})) for varying light-matter coupling strength $g$ (top curves) and driving strength $F$ (bottom curves) for Coulomb repulsion $U/t_{h} = 8$, cavity frequency $\Omega/t_{h} = 10$, and driving frequency $\Omega_{\text{dr}}/t_{h} = 10$. Grey dots above the curves indicate main peak positions. Curves are offset vertically for visibility.}
\label{fig:s}
\end{figure}
\begin{figure}[htp]
\centering
\includegraphics[width=0.5\textwidth]{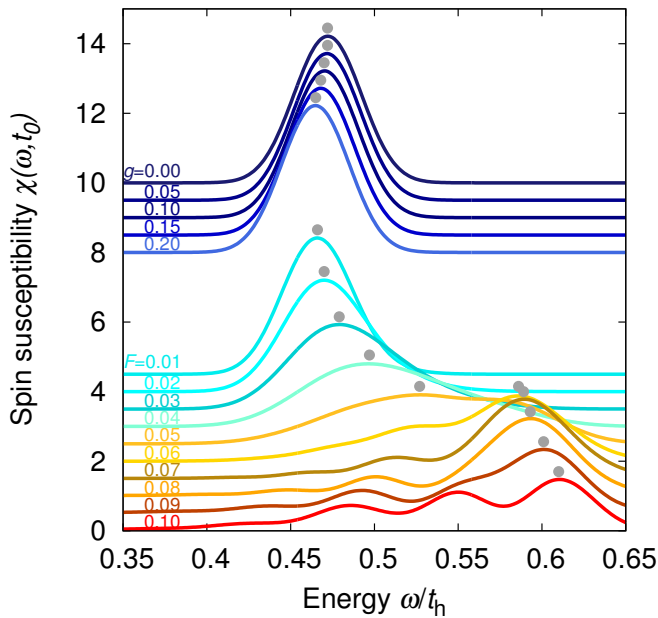}
\caption{{\bf Light-matter tuning of effective exchange interaction at sub-$U$ frequency.} Evolution of frequency-dependent spin susceptibility (Eq.~(\ref{eq:susc})) for varying light-matter coupling strength $g$ (top curves) and driving strength $F$ (bottom curves) for Coulomb repulsion $U/t_{h} = 8$, cavity frequency $\Omega/t_{h} = 6$, and driving frequency $\Omega_{\text{dr}}/t_{h} = 6$. Grey dots above the curves indicate main peak positions. Curves are offset vertically for visibility.}
\label{fig:s_sub}
\end{figure}

We first investigate the spin susceptibility as a function of the dimensionless light-matter coupling $g$. In practice, $g$ depends on the effective cavity volume and can be tuned by the specific cavity setup \cite{maissen_ultrastrong_2014,schlawin_cavity-mediated_2019}. Here we take it as a theory parameter to show the general effect of moderate light-matter coupling, which is realistically achievable. In Fig.~\ref{fig:s} the top blue-colored curves show the spin susceptibility as a function of energy $\omega$ for the undriven cavity and coupling strengths $g$ $=$ $0.00$ $\dots$ $0.20$ and a blue-detuned cavity frequency, equal to the driving frequency, $\Omega = 10$. Initially the peak position is $\omega$ $=$
0.472, which corresponds to the bare exchange coupling $J_{\text{eff}}$ and is slightly below the perturbative value $J_{\text{ex}}$ $=$ $0.5$.
When $g$ is increased, the peak moves to smaller values, e.g., $J_{\text{eff}} = 0.463$ for $g=0.20$.  

Starting from $g=0.20$ we then turn on the external driving field $F$. The evolution of the spin susceptibility under increasing $F$ shows that the peak position is further decreased. At the same time the curves broaden considerably. We note that a decrease of $J_{\text{eff}}$ for blue-detuned driving $\Omega > U$ is similarly obtained in the fully classical Floquet limit~\eqref{Jfl} \cite{mentink_ultrafast_2015}. 

To complement this behavior, we show in Fig.~\ref{fig:s_sub} the spin susceptibility evolving for sub-resonant, red-detuned frequency $\Omega = 6 < U$. First, for the dark cavity we observe again a reduction of $J_{\text{eff}}$, which is slightly less strong with $g$ compared to the blue-detuned case here. For instance, at $g=0.20$ we obtain $J_{\text{eff}} = 0.465$ compared to $0.463$ for the blue-detuned case. However, when the classical driving field is turned on, this reduction is quickly reversed and an enhancement of $J_{\text{eff}}$ is obtained, consistent with the analytical results discussed above in the paper. At larger driving fields, not only a broadening is found, but also the emergence of sidepeaks in the spin susceptibility. 

\begin{figure}[htp]
\centering
\includegraphics[width=0.5\textwidth]{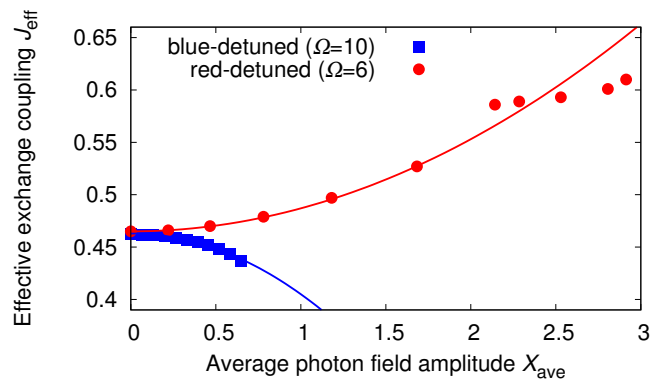}
\caption{Effective exchange coupling as extracted from the peak position in the local spin susceptibilities, corresponding to effective singlet-triplet splitting in the driven system, as a function of driving-induced average peak photon field amplitude, for blue and red detuning of the photon frequency as indicated. Data points correspond to driving field values $F$ $=$ $0.00$ $\dots$ $0.10$ in steps of $0.01$. Curves indicate quadratic scaling in the amplitude for small amplitudes. 
}
\label{fig:scaling}
\end{figure}

We summarize our findings for modifications of the effective exchange couplings in the driven cavity in Fig.~\ref{fig:scaling}, in which we show the extracted main peak positions as a function of the time-averaged photon field amplitude $X_{\text{ave}}$ $\equiv$  $\frac{1}{200}\int_0^{200} dt \; |X(t)|$. First of all, the maximally achieved amplitudes are larger for the same set of external field values $F$ in the red-detuned case, which is attributed to the different modifications of photon frequencies for $\Omega <U$ and $\Omega >U$ cases, see Eq.~\ref{phpononhgefak} and the appendix~\ref{appbbb} for details.  
Note that a similar dependence of $J_{\rm eff}$ on photon number $\langle \hat{a}^\dag\hat{a}\rangle$ instead of $X_{\rm ave}$ has also been found, which is expected because, as discussed in the analytical theory, a coherent displacement $\langle X\rangle\ne0$ is not needed to modify the exhange interaction. The reduction (enhancement) of $J_\text{eff}$ for blue (red) detuning is clearly visible here, and we find a quadratic dependence of this reduction (enhancement) on the driving-induced photon field amplitude, again consistent with the analytical results as well as the classical Floquet limit. 
 
However, a quantitative comparison is more difficult. First of all, the initial ground state features a mixture of different photon-number states instead of a a single one, so that the spin dynamics is not described by a single spin Hamiltonian, but is a result of contributions from all photon number sectors. More importantly, the classical driving itself further mixes different photon-number states during the time-evolution, averaging over the frequency peaks in the Fourier spectrum. This renders a quantitative analysis, though possible with the analytic theory (calculating the time-evolution using the spin-photon hamiltonian~\eqref{vghhssssh}), complicated and less relevant in this minimal model. Thus we reserve it for future studies with a more realistic cavity setup.

{\color{black} Finally, we comment on the experimentally relevant effects of decoherence and dissipation in the case of a driven cavity, which are not included in the our simulations of a closed electron-photon system. First of all, one should state that these effects will result mainly from coupling to other degrees of freedom in the material under consideration, such as phonons or a substrate
, and the resulting thermalization of the electronic system may stimulate photon absorption from the driven cavity.
 Cavity losses can also play a role but usually occur on longer time scales. It is clear that such decoherence and dissipation effects might limit the time scales on which cavity-modified exchange interactions will be achievable in practice. On the other hand, one of our central results is that coherence in the photon states is not a necessary prerequisite to achieve such cavity-modified interactions, provided that relatively strong light-matter coupling can be achieved. Therefore we expect the limitations of decoherence to be more severe at weak light-matter coupling than at strong light-matter coupling.}

\section{Conclusion and Outlook}
\label{Sec:conclusion}
In this work, we have investigated the light-induced changes in material properties from the quantum to the classical limit. 
In general, the classical limit is achieved by increasing the photon number $n$, while decreasing the coupling, so that $g\sqrt{n}$ is fixed. 
We have introduced a cavity Schrieffer-Wolff transformation to derive the spin-exchange interaction in the presence of quantum light coupling. In particular, we observed that the cavity-modified spin-exchange interaction deviates from the bare value $J_{\rm ex}=4t_h^2/U$ and matches both the Floquet result \cite{mentink_ultrafast_2015} $J^F_{\rm ex}$ in the classical limit, and the vacuum renormalization of $J_{ex}$ \cite{kiffner_manipulating_2019} for the dark cavity. By systematical examination of the quantum-light regime, we show that the Floquet engineering can be extended to the few-photon regime.
In particular, already putting the cavity in the one-photon state is enough to revert the sign of $J_{ex}$, which is possible in the Floquet limit but not for the dark cavity. 

A coherent amplitude of the photon field is not needed to recover the Floquet physics. Instead, if the cavity is in a coherent state with few photons, the dynamics of the matter is not described by a single effective Hamiltonian but instead shows individual contributions from each photon number sector. In the classical limit of weak coupling and macroscopically large photon number, on the other hand, both the photon number state and the coherent state, and in fact any photon occupation sharply centered around some large average photon number $\bar{N}$, recovers the Floquet limit. As a result, the assumption of a coherent driving is apparently too strong for the purpose of Floquet engineering.
{\color{black}
The slightly counterintuitive observation may be clarified by the fact that the effective exchange interaction in  a classical coherent state is also obtained by virtual emission and absorption of photons. The phase information between the different Fock state components of the cavity state is thereby lost because a phase $e^{i\phi}$ on emission is cancelled by a corresponding phase $e^{-i\phi}$ on absorption.
}


Moreover, we studied a minimal numerical model,
 where the cavity is driven by a classical field. In this minimal setup, the cavity acts as a transducer from the external driving laser to the electrons in a material. 
 A photon amplitude is created by the external driving, and then drives the electronic system to modify its the spin susceptibility. The cavity-induced modification observed in the simulation is fully consistent with the analytic solution on the coherent state and photon-number state. It is shown that the Floquet-like engineering is indeed restored with a relatively weak photon amplitude, in the strong light-matter coupling regime. In contrast to the Floquet limit where excess heating is often unavoidable, a state with finitely many photons can minimize the heating effect, due to the limited electromagnetic energy inside the cavity volume. Our findings therefore open up new design opportunities for electronic devices with high tunability, low energy consumption, and minimal heating effects.


On the fundamental side, it will be also intriguing to further investigate how the control of induced interactions in solids by coupling to driven cavities can tune properties of quantum materials. In the future, it would be promising to extend the current results to a lattice model to examine the nature of mixed photon-magnon excitations. For example, the study of cavity-modified spin fluctuations and their influence on high-temperature superconductivity promises intriguing insights \cite{dahm_strength_2009,keimer_quantum_2015}. Using the cavity-Schrieffer Wolff transformation, it will also be interesting to investigate the Kondo physics and generally quantum criticality in the cavity-coupled system, as well as to look for cavity realizations of light-induced scalar spin chirality proposals \cite{claassen_dynamical_2017,kitamura_probing_2017,owerre_floquet_2017,elyasi_topologically_2019} with time-reversal symmetry-breaking photon fields \cite{wang_cavity_2019}. Another direction would be to study more realistic models, especially to include the effect of an open cavity, where a non-equilibrium steady-state of the light-matter system can be maintained through weak external driving. 

{\color{black} Moreover, it will be interesting to consider long-range interactions induced via the cavity. In Ref.~\cite{kiffner_manipulating_2019}, such terms appear in form of hopping processes on two distant bonds which become correlated via virtual photon exchange. While such terms are not present in the spin model, where all charge excitations have been projected out, in the spin model one can expect corresponding correlated spin flips on distant bonds. To systematically derive such terms, one would have to go to fourth order perturbation theory in the hopping, which is possible, but left for future work. A subset  of these terms, proportional to $t_h^4/(U^2\Omega)$, can be obtained by perturbatively eliminating the two-photon creation and annihiliation processes from the  spin-photon Hamiltonian. Though such terms are smaller in $t_h^2/U$, their long-range character might make them highly relevant for the resulting spin models, in particular as long range spin interactions can give rise to frustration.}

The discussion of cavity-Floquet crossover can also be extended to more general contexts, such as in the intermediate or weak interaction regime. Indeed, without performing the Schrieffer-Wolff transformation, the light-matter Hamiltonian~\eqref{hgejde2kjh2ekln} represented on the photon-number basis constitutes a natural analogy of the Floquet Hamiltonian, and a general light-matter system is expected to show a similar cavity-Floquet crossover, which can make a bridge between general Floquet-engineered states and cavity states.  

\section*{ACKNOWLEDGEMENT} 
M.A.S.~acknowledges financial support by the DFG through the Emmy Noether program (SE 2558/2-1).
J.L. and M.E.~were supported by the ERC starting grant No. 716648.

\bibliography{Floquet_Crossover,Floquet_Crossover_additional}

\onecolumngrid
\appendix

\section{Derivation}
\label{Sec:dervation}

To bring the Hamiltonian into some suitable form, we use time-dependent unitary transformations. 
For a general unitary transformation $W(t)$,  we define the transformation to the rotating frame as 
$|\psi_{rot}(t)\rangle=W(t)|\psi(t)\rangle$. The new wave-function satisfies Schr\"odinger equation 
$i\partial_t |\psi_{rot}(t)\rangle=H_{rot}(t) |\psi_{rot}(t)\rangle$ with
\begin{align}
H_{rot}(t)
=
W(t)
[H
-
i\partial_t
]
W(t)^\dagger.
\end{align}
We first use a basis rotation to remove the free photon Hamiltonian from \eqref{hgejde2kjh2ekln}.
With $W(t)=e^{it\Omega  a^\dagger a  }$ we have 
\begin{align}
H_{rot}(t)&= t_{h} \sum_\sigma \left( \hat{c}_{1,\sigma}^\dagger \hat{c}_{2,\sigma}^{} \; e^{i\hat{A}(t)} + \textrm{H.c.} \right) + U\hat D
\,
\equiv
\, \alpha \hat T(t)+ \hat V,
    \label{hgejde2kjh2ekln01}
  \end{align}
where 
\begin{align}
\hat A(t)=W(t)\hat A W^\dagger(t)
=
a e^{-i\Omega t} + a^\dagger e^{i\Omega t},
 \end{align}
 and the dimensionless parameter has been inserted as an expansion parameter ($\alpha\ll1$.)

\subsection{Dynamical localization}
\label{sec:appA1}
Note that the Hamiltonian~\eqref{hgejde2kjh2ekln01} is now periodic in time. To understand its high-frequency limit, we perform the (Van-Vleck) high-frequency expansion and only retain the zeroth order term, which is the time-average of the Hamiltonian over a period $2\pi/\Omega$,

\begin{align}
H_{\rm eff}&=\frac{\Omega}{2\pi}\int_0^{\frac{2\pi}{\Omega}}dtH_{rot}(t)\nonumber\\
&= t_{h} \sum_\sigma \left( \hat{c}_{1,\sigma}^\dagger \hat{c}_{2,\sigma}^{} \; \int^1_0 dx e^{ig(a e^{-2\pi ix}+a^\dag e^{2\pi ix})} + \textrm{H.c.} \right) + U\hat D
\,
\equiv
\, \alpha \hat T(t)+ \hat V,
\end{align}
which can be evaluated by a Taylor expansion of the exponential $e^{ig(a e^{-2\pi ix}+a^\dag e^{2\pi ix})}$. In fact, we have
\begin{align}
\int^1_0 dx e^{ig(a e^{-2\pi ix}+a^\dag e^{2\pi ix})}&=\int^1_0 dx e^{iga^\dag e^{2\pi ix}}e^{iga e^{-2\pi ix}}e^{-g^2/2}\nonumber\\
&=e^{-g^2/2} \sum_{kk'}\frac{(iga^\dag)^k}{k!}\frac{(iga)^{k'}}{k'!}\int^1_0 dx e^{2\pi i(k-k')x}\nonumber\\
&=e^{-g^2/2} \sum_{kk'}\frac{(iga^\dag)^k}{k!}\frac{(iga)^{k'}}{k'!}\delta_{k,k'}\nonumber\\
&=e^{-g^2/2} \sum_{k}\frac{(ig)^{2k}{a^\dag}^ka^k}{k!^2},
\end{align}
which is nothing but the $\hat{J}_{h}(2g)$ defined in the main text.

To take the large photon number limit, we note that
\begin{align}
J^{(n)}_{h}(2g)=\langle n|\hat{J}_{h}(2g)|n\rangle&=e^{-g^2/2} \sum_{k=0}^n\frac{(ig)^{2k}}{k!^2}\frac{n!}{(n-k)!},
\end{align}
which is a finite sum that can be readily evaluated. In the limit $n\to\infty$ with $2g\sqrt{n}=A$, we have $n!/(n-k)!\to n^k$, and 
\begin{align}
J^{(n)}_{h}(2g)\to e^{-A^2/8n} \sum_{k}\frac{(-1)^k(A/2)^{2k}}{k!^2}\to J_0(A).
\end{align}
This is the Floquet result.

\subsection{Schrieffer-Wolff transformation}
In the next step, we attempt a time-dependent unitary transformation $W_2(t)=e^{S(t)}$, which is designed to make the Hamiltonian diagonal in the double occupancy, in order to facilitate a projection to the spin sector: We define projectors $\mathcal{P}_0$ and $\mathcal{P}_1=1-\mathcal{P}_0$ to sectors $0$ and $1$, and decompose each operator $A$ into transitions $A_{ab}\equiv \mathcal{P}_a A \mathcal{P}_b$ between and within the sectors. We attempt to find  a  time-dependent unitary transformation  $W_2(t)=e^{S(t)}$ (parametrized by the antihermitian matrix $S$), such that in the rotating basis matrix elements between sectors $0$ and $1$ vanish {\em at any time} \cite{eckstein_designing_2017}. A Taylor ansatz $S=\alpha S_1 + \alpha^2 S_2 + \cdots$ yields the series 
\begin{align}
&H_{rot}(t)
=
 V 
+
\alpha 
\big\{
 T
+
[S_1, V]
+
i\dot S_1
\big\}
+
\alpha ^2
\big\{
[S_2, V]
+
i \dot S_2
+
[S_1, T]
+
\frac{1}{2}
[S_1,i\dot S_1 + [S_1, V]]
\big\}
+
\mathcal{O}(\alpha^3).
\label{hrotexpansion}
\end{align}
One can now truncate the expansion of $S$ after a given order $m$, and choose $S_m$ such that $H_{rot}$ has no mixing terms up to  order $m$. Here we proceed even simpler, looking for a time-periodic solution for the generator $S(t)$. We request that the first order has no transition matrix elements
\begin{align}
\label{bshqhjxaa}
 T_{01}+T_{10}
+
[S_1, V]
+
i\dot S_1
=0.
\end{align}
Since all operators are periodic with period $T=2\pi/\omega$, we can use a Fourier decomposition 
\begin{align}
A(t)&=\sum_{n} A^{(n)}e^{-i\omega nt} 
\\
A^{(n)}&=
\frac{1}{T}
\int_{0}^T dt \,
A(t) e^{i\omega n t}.
\end{align}
With the ansatz $S_1\equiv S_{10}+S_{01}$, Eq.~\eqref{bshqhjxaa} becomes
\begin{align}
\label{order1}
&0
=
 T_{01}^{(n)}
 +
 T_{10}^{(n)}
+
[S_{01}^{(n)} + S_{10}^{(n)}, V]
+
n\omega
(S_{01}^{(n)} + S_{10}^{(n)})
\end{align}
Since $V_{00}=0$, 
\begin{align}
\nonumber
0
&=
 T_{01}^{(n)}
+
S_{01}^{(n)} V_{11}
+
n\omega
S_{01}^{(n)}
\\
0
&=
 T_{10}^{(n)}
-
V_{11}S_{10}^{(n)} 
+
n\omega
S_{10}^{(n)},
\end{align}
and thus
\begin{align}
\nonumber
S_{01}^{(n)} &=
- T_{01}^{(n)}[U+n\omega]^{-1}
\\
S_{10}^{(n)} &=
[U-n\omega]^{-1}  T_{10}^{(n)}.
\end{align}
Using Eq.~\eqref{bshqhjxaa} in Eq.\eqref{hrotexpansion}, we obtain, for the second order terms
\begin{align}
\alpha^2\Big\{
[S_2,V]+i\partial_t S_2 +  [S_1,T_{11}] + \tfrac12[S_1,T_{01}+T_{10}]
\Big\}. 
\end{align}
Proceeding as before, all second order terms which mix sector $0$ and $1$ of the Hilbert space, such as, e.g., the gerenated terms $S_{01}T_{11}$, are removed by a choice of $S_2$. The terms which remain in sector $0$ are from the last commutator,
\begin{align}
\frac{1}{2}[S_{01}T_{10} - T_{01}S_{10}].
\end{align}
Inserting Fourier components, the Hamiltonian in the $00$ sector is 
\begin{align}
H^{(n)}_{rot}&=
\frac{\alpha^2}{2}
\sum_{l}\big[  S_{01}^{(n-l)}T_{10}^{(l)} - T_{01}^{(n-l)}S_{10}^{(l)}\big]
\\
&=
-\frac{\alpha^2}{2}
\sum_{l}\Big[  
\frac{ T_{01}^{(n-l)}T_{10}^{(l)}}
{U+(n-l)\omega}
 +
\frac{
T_{01}^{(n-l)}T_{10}^{(l)}}
{U-l\omega}  
\Big].
\\
&=
-\frac{\alpha^2}{2}
\sum_{m,l}\delta_{m+l,n}\Big[  
\frac{
T_{01}^{(m)}T_{10}^{(l)}}
{U+m\omega}  
+
\frac{ T_{01}^{(m)}T_{10}^{(l)}}
{U-l\omega}
\label{gkkehjke}
\Big].
\end{align}
We now evaluate the time-dependent operators. For this, it is convenient to 
intrododuce 
\begin{align}
B^{(m)}_g
&=
\frac{1}{T}
\int_0^{T} dt \,e^{igA(t)}e^{im \omega t}.
\end{align}
Then, for a bond $(rs)$,
\begin{align}
T_{01}^{(m)} =
t_{h}\sum_{\sigma} [c_{r\sigma}^\dagger c_{s\sigma}]_{01} B^{(m)}_g +  [c_{s\sigma}^\dagger c_{r\sigma}]_{01} B^{(m)}_{-g},
\end{align}
and
\begin{align}
T_{01}^{(m)} T_{10}^{(l)}
&=
t_{h}^2\sum_{\sigma}\Big[ [c_{r\sigma}^\dagger c_{s\sigma}]_{01} B^{(m)}_g +  [c_{s\sigma}^\dagger c_{r\sigma}]_{01} B^{(m)}_{-g}\Big]
\sum_{\sigma'}\Big[ [c_{r\sigma'}^\dagger c_{s\sigma'}]_{10} B^{(l)}_g +  [c_{s\sigma'}^\dagger c_{r\sigma'}]_{10} B^{(l)}_{-g}\Big]
\\
&=
t_{h}^2
\sum_{\sigma,\sigma'}\Big[ 
[c_{r\sigma}^\dagger c_{s\sigma}]_{01}  [c_{s\sigma'}^\dagger c_{r\sigma'}]_{10} B^{(m)}_g B^{(l)}_{-g}
+
 [c_{s\sigma}^\dagger c_{r\sigma}]_{01}  [c_{r\sigma'}^\dagger c_{s\sigma'}]_{10} B^{(m)}_{-g}B^{(l)}_g
 \label{hefeehe}
\Big].
\end{align}
The projected hoppings reduce to spin operators in the $00$ sector as usual,
\begin{align}
&[c_{r\sigma}^\dagger c_{s\sigma}]_{01}  [c_{s\sigma'}^\dagger c_{r\sigma'}]_{10}
=
c_{r\sigma}^\dagger(1-n_{r\bar\sigma}) c_{s\sigma}n_{s\bar\sigma}  c_{s\sigma'}^\dagger n_{s\bar\sigma'} c_{r\sigma'}(1-n_{r\bar\sigma'})
\\
&=
\delta_{\sigma,\sigma'}
\Big[c_{r\sigma}^\dagger(1-n_{r\bar\sigma}) c_{s\sigma}n_{s\bar\sigma}  c_{s\sigma}^\dagger n_{s\bar\sigma} c_{r\sigma}(1-n_{r\bar\sigma})
\Big]
+
\delta_{\bar\sigma,\sigma'}
\Big[c_{r\sigma}^\dagger(1-n_{r\bar\sigma}) c_{s\sigma}n_{s\bar\sigma}  c_{s\bar\sigma}^\dagger n_{s\sigma} c_{r\bar\sigma}(1-n_{r\sigma})
\Big]
\\
&=
\delta_{\sigma,\sigma'}
(1-n_{r\bar\sigma}) n_{r\sigma} 
(1-n_{s\sigma}) n_{s\bar\sigma} 
+
\delta_{\bar\sigma,\sigma'}
\Big[
c_{r\sigma}^\dagger
c_{r\bar\sigma}
(1-n_{r\sigma})
c_{s\sigma}
c_{s\bar\sigma}^\dagger 
n_{s\sigma} 
\Big],
\end{align}
so that, after projection to the $00$ sector,
\begin{align}
\sum_{\sigma,\sigma'}
[c_{r\sigma}^\dagger c_{s\sigma}]_{01}  [c_{s\sigma'}^\dagger c_{r\sigma'}]_{10}
=
\frac{1}{2}- 2 S_{r}^z S_{s}^z
-
S_{r}^+ S_{s}^--S_{r}^- S_{s}^+
=
\frac{1}{2}- 2 \bm S_{r} \bm S_{s}
\equiv
2\mathcal{P}^s_{rs}.
\end{align}
($\mathcal{P}^s_{rs}$ is actually the projector to the singlet on bond $rs$).
Using this expression in \eqref{hefeehe} we have
\begin{align}
T_{01}^{(m)} T_{10}^{(l)}
&=
2t_{h}^2\mathcal{P}^s_{rs}
\Big(B^{(m)}_g B^{(l)}_{-g}+B^{(m)}_{-g} B^{(l)}_{g}\Big).
\end{align}
The full exchange Hamiltonian is, using \eqref{gkkehjke},
\begin{align}
H^{(n)}_{rot}
&=
-\alpha^2 \mathcal{P}^s_{rs}
\sum_{m,l}\delta_{m+l,n}
\Big(B^{(m)}_g B^{(l)}_{-g}+B^{(m)}_{-g} B^{(l)}_{g}\Big)
\Big(  \frac{t_{h}^2}{U-l\omega} +\frac{t_{h}^2}{U+m\omega}  \Big),
\end{align}
and thus
\begin{align}
H_{rot}(t)&=\sum_{n}H^{(n)}_{rot} e^{-in\omega t}
=
-\alpha^2 \mathcal{P}^s_{rs}\mathcal{J}(t)
\\
\mathcal{J}(t)
&=
\sum_{n,l}
e^{-in\omega t}
\Big(B^{(n-l)}_g B^{(l)}_{-g}+B^{(n-l)}_{-g} B^{(l)}_{g}\Big)
\Big(  \frac{t_{h}^2}{U-l\omega} +\frac{t_{h}^2}{U+(n-l)\omega}  \Big).
\label{ggecefee}
\end{align}
We now evaluate the exchange operator $\mathcal{J}(t)$. First, consider
\begin{align}
B^{(n-l)}_g B^{(l)}_{-g}=
\int\frac{dt_1dt_2}{T^2}
e^{igA(t_1)}e^{-igA(t_2)}e^{i\omega[(n-l)t_1+lt_2]}.
\label{gegegeeh}
\end{align}
Using the Baker-Hausdorff relation
(where $[X,Y]$ is a c-number),
\begin{align}
e^{X+Y}&=e^{X}e^{Y}e^{-[X,Y]/2},
\\
e^{X}e^{Y} 
&=
e^{Y}e^{X} 
e^{[X,Y]},
\end{align}
we can normal order the integrand with respect to $a$ and $a^\dagger$, 
\begin{align}
&e^{igA(t_1)}
e^{-igA(t_2)}
=
e^{igae^{-i\omega t_1} +iga ^\dagger e^{i\omega t_1} }
e^{-igae^{-i\omega t_2} -iga ^\dagger e^{i\omega t_2} }
\\
&=
e^{iga ^\dagger e^{i\omega t_1} }
e^{igae^{-i\omega t_1}}
e^{-[iga ^\dagger e^{i\omega t_1} ,igae^{-i\omega t_1}]/2}
e^{ -iga ^\dagger e^{i\omega t_2}}
e^{-igae^{-i\omega t_2}}
e^{-[ -iga ^\dagger e^{i\omega t_2},-igae^{-i\omega t_2}]/2}
\\
&=
e^{iga ^\dagger e^{i\omega t_1} }
e^{igae^{-i\omega t_1}}
e^{ -iga ^\dagger e^{i\omega t_2}}
e^{-igae^{-i\omega t_2}}
e^{-g^2}
\\
&=
e^{iga ^\dagger e^{i\omega t_1} }
\,
e^{ -iga ^\dagger e^{i\omega t_2}}
e^{igae^{-i\omega t_1}}
e^{[igae^{-i\omega t_1}, -iga ^\dagger e^{i\omega t_2}]}
\,
e^{-igae^{-i\omega t_2}}
e^{-g^2}
\\
&=
e^{iga ^\dagger (e^{i\omega t_1} -e^{i\omega t_2})}
e^{iga(e^{-i\omega t_1}-e^{-i\omega t_2})}
e^{g^2(e^{i\omega (t_2-t_1)}-1) }
\\
&=
e^{iga ^\dagger e^{i\omega t_1} (1 -e^{i\omega (t_2-t_1)})}
e^{igae^{-i\omega t_1}(1-e^{-i\omega (t_2-t_1)})}
e^{g^2(e^{i\omega (t_2-t_1)}-1) }.
\end{align}
Using $a(t)=ae^{-i\omega t}$,  and $t_2-t_1\equiv t_r$,
\begin{align}
&e^{igA(t_1)}
e^{-igA(t_2)}
=
e^{ig a(t_1)^\dagger  (1 -e^{i\omega t_r})}
e^{iga(t_1)(1-e^{-i\omega t_r})}
e^{g^2(e^{i\omega t_r}-1) }.
\end{align}
Inserted into \eqref{gegegeeh}
\begin{align}
e^{-in\omega t}
B^{(n-l)}_g B^{(l)}_{-g}=
\int\frac{dt_1dt_2}{T^2}
e^{ig a^\dagger(t_1)  (1 -e^{i\omega t_r})}
e^{iga(t_1)(1-e^{-i\omega t_r})}
e^{g^2(e^{i\omega t_r}-1) }
e^{i\omega n(t_1-t)}
e^{i\omega l(t_2-t_1)}.
\label{gegegeeh01}
\end{align}
We can now add back the $\omega a^\dagger a$ term to the Hamiltonian (inverse of the first unitary rotation $W$), which corresponds to a shift of the time-arguments $t_1\to t_1-t$ in the operators $a$ and $a^\dagger$. With a shift of the integration variable $ t_1-t\to s$, \eqref{gegegeeh01} gives
\begin{align}
e^{-in\omega t}
B^{(n-l)}_g B^{(l)}_{-g}&=
\int\frac{ds dt_r}{T^2}
e^{ig a(s)^\dagger  (1 -e^{i\omega t_r})}
e^{iga(s)(1-e^{-i\omega t_r})}
e^{g^2(e^{i\omega t_r}-1) }
e^{i\omega ns}
e^{i\omega l t_r}
\end{align}
Taylor expansion of the exponentials,
\begin{align}
e^{-in\omega t}
B^{(n-l)}_g B^{(l)}_{-g}&=
\sum_{b,c=0}^\infty
\int\frac{ds dt_r}{T^2}
\frac{[ig a(s)^\dagger  (1 -e^{i\omega t_r})]^b}{b!}
\frac{[iga(s)(1-e^{-i\omega t_r})]^c}{c!}
e^{g^2(e^{i\omega t_r}-1) }
e^{i\omega ns}
e^{i\omega l t_r}.
\end{align}
Now one can see that the $s$-integral projects to $n+b-c=0$,
\begin{align}
e^{-in\omega t}
B^{(n-l)}_g B^{(l)}_{-g}&=
\sum_{b,c=0}^\infty
\delta_{n,c-b}
(a^\dagger)^ba^c
\int\frac{dt_r}{T}
\frac{[ig   (1 -e^{i\omega t_r})]^b}{b!}
\frac{[ig(1-e^{-i\omega t_r})]^c}{c!}
e^{g^2(e^{i\omega t_r}-1) }
e^{i\omega l t_r}.
\label{khe3xblwnjbqk}
\end{align}
In Eq.~\eqref{ggecefee}  one must add the term \eqref{khe3xblwnjbqk} and a corresponding term with  $g\to-g$, which corresponds to a multiplication of \eqref{khe3xblwnjbqk}  with $(1+(-1)^{b+c})$,
\begin{align}
\mathcal{J}
&=
\sum_{l,n}
\sum_{b,c=0}^\infty
\delta_{n,b-c}
(a^\dagger)^ba^c
[1+(-1)^{n}]
\Big(  \frac{t_{h}^2}{U-l\omega} +\frac{t_{h}^2}{U-(n+l)\omega}  \Big)\,\,\,\times
\\
&\times\,\,\,\,\,\,\int_0^1 dx
\frac{[ig   (1 -e^{i2\pi x})]^b}{b!}
\frac{[ig(1-e^{-i2\pi x})]^c}{c!}
e^{g^2(e^{i2\pi x}-1) }
e^{i2\pi x l},
\\
&=
\sum_{l,n}
\sum_{b,c=0}^\infty
\delta_{n,b-c}
(a^\dagger)^ba^c
[1+(-1)^{n}]
\Big(  \frac{t_{h}^2}{U-l\omega} +\frac{t_{h}^2}{U-(n+l)\omega}  \Big)\,\,\,\times
\\
&\times\,\,\,\,\,\,\int_0^1 dx
\frac{[ig   (e^{-i\pi x} -e^{i\pi x})]^b}{b!}
\frac{[ig(e^{i\pi x}-e^{-i\pi x})]^c}{c!}
e^{g^2(e^{i2\pi x}-1) }
e^{i2\pi x l}e^{i\pi x n} ,
\end{align}
where we have changed $n\to-n$ for convenience and substituted $\omega t_r=2\pi x$.
The number $-n$ counts the change in the  photon number. It is therefore useful to represent
\begin{align}
\mathcal{J}=\mathcal{J}_0+ 
\sum_{n=2,4,..} \big[ (a^\dagger)^n\mathcal{J}_n + \mathcal{J}_{-n}a^n],
\end{align}
where $\mathcal{J}_n$ are operators which are diagonal in the photon number. 
For the terms with $n\ge0$ we get ($n=2m$)
\begin{align}
\mathcal{J}_{n}
&=
\sum_{l}
\sum_{c=0}^\infty
(a^\dagger)^{c}a^c
\Big(\frac{2t_{h}^2}{U-l\omega} +\frac{2t_{h}^2}{U-(n+l)\omega}  \Big)
\int_0^1 dx
\frac{[g (e^{i\pi x} -e^{-i\pi x})]^{2(c+m)}}{(c+2m)!c!}
e^{g^2(e^{i2\pi x}-1) }
e^{i2\pi x (l+m)}
\\
&=
\sum_{l}
\sum_{c=0}^\infty
(a^\dagger)^{c}a^c
\Big(\frac{2t_{h}^2}{U-(l-m)\omega} +\frac{2t_{h}^2}{U-(l+m)\omega}  \Big)
\int_0^1 dx
\frac{[g (e^{i\pi x} -e^{-i\pi x})]^{2(c+m)}}{(c+2m)!c!}
e^{g^2(e^{i2\pi x}-1) }
e^{i2\pi x l}.
\end{align}
One can see that  $\mathcal{J}_{n}$ is hermitian, and with some math, that $\mathcal{J}_{-n}=\mathcal{J}_{n}$ (so that $\mathcal{J}$ is hermitian). 

To explicitly evaluate the expressions, we expand the product,
\begin{align}
\mathcal{J}_{2m}
&=
e^{-g^2}
\sum_{l}
\sum_{c=0}^\infty
g^{2(c+m)}
\frac{(a^\dagger)^{c}a^c}{(c+2m)!c!}
\Big(\frac{2t_{h}^2}{U-l\omega} +\frac{2t_{h}^2}{U-(2m+l)\omega}  \Big)
\sum_{p=0}^{2(c+m)}
(-1)^p\begin{pmatrix}
2c+2m
\\
p
\end{pmatrix}
\int_0^1 dx
 e^{-i2\pi x(c-p-l)}
e^{g^2e^{i2\pi x} }\nonumber
\\
&=
e^{-g^2}
\sum_{l}
\Big(\frac{2t_{h}^2}{U-l\omega} +\frac{2t_{h}^2}{U-(2m+l)\omega}  \Big)
\sum_{c=0}^\infty
(a^\dagger)^{c}a^c
\frac{g^{2(c+m)}}{(2c+2m)!}
\begin{pmatrix} 2c+2m\\c\end{pmatrix} \times
\nonumber \\
& \hspace{1cm} \times \sum_{p=0}^{2(c+m)}
(-1)^p
\begin{pmatrix} 2c+2m\\p\end{pmatrix}
\int_0^1 dx
 e^{-i2\pi x(c-p-l)}
e^{g^2e^{i2\pi x} }.\nonumber
\end{align}
The integral evaluates to 
\begin{align}
\int_0^1 dx
 e^{-i2\pi x(c-p-l)}e^{g^2e^{i2\pi x} }
 =
 \sum_{r=0}^\infty
 \frac{g^{2r}}{r!}
 \delta_{c-p-l,r}.
\end{align}
Using this result, we get
\begin{align}
\mathcal{J}_{2m}
&=
J_{\rm ex}
\frac{e^{-g^2}}{2}\sum_{l}
\Big(\frac{1}{1-l\bar\omega} +\frac{1}{1-(2m+l)\bar\omega}  \Big)
\sum_{c=0}^\infty
(a^\dagger)^{c}a^c g^{2c}\,
\,\,\,\,\,\times 
\\
&\times\,\,\,\,
\frac{g^{2m}}{(2c+2m)!}
\begin{pmatrix} 2c+2m\\c\end{pmatrix}
 \sum_{r=0}^\infty
 \frac{g^{2r}}{r!}
\sum_{p=0}^{2(c+m)}
(-1)^p
\begin{pmatrix} 2c+2m\\p\end{pmatrix}
 \delta_{c-p-l,r}
\\
&=
\frac{J_{\rm ex}}{2}
\sum_{c=0}^\infty
(a^\dagger)^{c}a^c\,
\frac{e^{-g^2}g^{2m+2c}}{(2c+2m)!}
\begin{pmatrix} 2c+2m\\c\end{pmatrix}
\,\,\,\,\,\times 
\\
&\times\,\,\,\,
\sum_{r=0}^\infty
\frac{g^{2r}}{r!}
\sum_{p=0}^{2(c+m)}
(-1)^p
\begin{pmatrix} 2c+2m\\p\end{pmatrix}
\Big(
\frac{1}{1+(r-c+p)\bar\omega} +\frac{1}{1+(r-c+p-2m)\bar\omega}  
\Big)
\end{align}
By defining
\begin{align}
L_p(\bar\omega,g)=
e^{-g^2}
\sum_{r=0}^\infty
\frac{g^{2r}}{r!}
\frac{1}{1+(r+p)\bar\omega},
\end{align}
and
\begin{align}
\mathcal{L}_{c,m}(\bar\omega,g)
&=
\frac{1}{2(c+2m)!c!}
\sum_{p=0}^{2(c+m)}
(-1)^p
\begin{pmatrix} 2c+2m\\p\end{pmatrix}
\big(L_{p-c}(\bar\omega,g)+L_{p-c-2m}(\bar\omega,g)\big)
\label{specialfunction}
\end{align}
we finally get Eq.~\eqref{debwelwe}.

\subsection{Details on the classical limit}
The exchange coupling $J^{(n)}_{\rm ex}=\langle n|\mathcal{J}_0|n\rangle$ can be evaluated to be
\begin{align}
\nonumber\\
J_{\rm ex}^{(n)}=
J_{\rm ex}
\sum_{c=0}^n
\frac{g^{2c}n!}{(n-c)!c!c!}\,
\sum_{p=0}^{2c}
\begin{pmatrix} 2c\\p\end{pmatrix}
(-1)^p
L_{p-c}(\bar\omega,g),
\label{gwgeheeee}
\end{align}
where we used that $(a^\dagger)^c a^c |n\rangle = \frac{n!}{(n-c)!}|n\rangle$ if $n\ge c$. The expression \eqref{gwgeheeee} is a finite double sum which is readily evaluated. 

Recall the explicit form of the Bessel function,
\begin{align}
J_{|\ell|}(A)=\sum_{p=0}^\infty\frac{(-1)^p}{p!(|\ell|+p)!}(A/2)^{2p+|\ell|}
\end{align}
Under the limit $n\to\infty$ and $2g\sqrt{n}=A$, one again notes that $n!/(n-c)!\to n^c$, and thus,
\begin{align}
J_{\rm ex}^{(n)}
&\to
J_{\rm ex}
\sum_{c=0}^\infty
\frac{(A/2)^{2c}}{c!^2}
\sum_{p=0}^{2c}
\begin{pmatrix} 2c\\p\end{pmatrix}
\frac{(-1)^p}{1+(p-c)\bar\omega}
\nonumber\\
&=
J_{\rm ex}
\sum_{l}
\frac{1}{1-l\bar\omega}
\sum_{c=0}^\infty
\frac{(A/2)^{2c}}{c!^2}
\sum_{p=0}^{2c}
\begin{pmatrix} 2c\\p\end{pmatrix}
(-1)^p\delta_{c-p,l}
\\
&=
J_{\rm ex}
\sum_{l}
\frac{1}{1-\ell\bar\omega}
\sum_{p=0}^{\infty}
\frac{(-1)^p(A/2)^{2(p+|\ell|)}}{(p+|\ell|)!^2}
\begin{pmatrix} 2(p+|\ell|)\\p\end{pmatrix}.
\label{bshjswhwhw}
 \end{align}
Using the Bessel function
\begin{align}
J_{|\ell|}(A)=\sum_{p=0}^\infty\frac{(-1)^p}{p!(|\ell|+p)!}(A/2)^{2p+|\ell|},
\end{align}
one can verify that 
\begin{align}
J_{|\ell|}(A)^2&=\sum_{kk'}\frac{(-1)^{k+k'}\left(\frac{A}{2}\right)^{2(k+k')+2|\ell|}}{k!k'!(|\ell|+k)!(|\ell|+k')!}\nonumber\\
&=\sum_{p=0}^\infty(-1)^p\left(\frac{A}{2}\right)^{2p+2|\ell|}\sum_{k=0}^{p}\frac{1}{k!(p-k)!(|\ell|+k)!(p+|\ell|-k)!}\nonumber\\
&=\sum_{p=0}^\infty\frac{(-1)^p}{p!(p+2|\ell|)!}\left(\frac{A}{2}\right)^{2p+2|\ell|}\sum_{k=0}^{p}{p\choose k}{p+2|\ell|\choose k+|\ell|}\nonumber\\
&=\sum_{p=0}^\infty\frac{(-1)^p}{(p+|\ell|)!^2}\left(\frac{A}{2}\right)^{2p+2|\ell|}{2p+2|\ell|\choose p},
\end{align}
which implies \eqref{Jfl}. Note that for the third equality we have used the following identity (for $k'\ge0,n+k'\le m$):
\begin{align}
\label{small}
\sum_{k=0}^n{n \choose k}{m \choose k+k'}={n+m \choose m-k'}.
\end{align}
It can be checked by comparing the constant term ($x^{-k'}$) from the LHS and RHS of the identity $(1+x)^n(1+1/x)^m=(1+x)^{n+m}/x^m$.

\section{Photon displacements and power spectra}
\label{appbbb}
\begin{figure*}[htp]
\centering
\includegraphics[width=\textwidth]{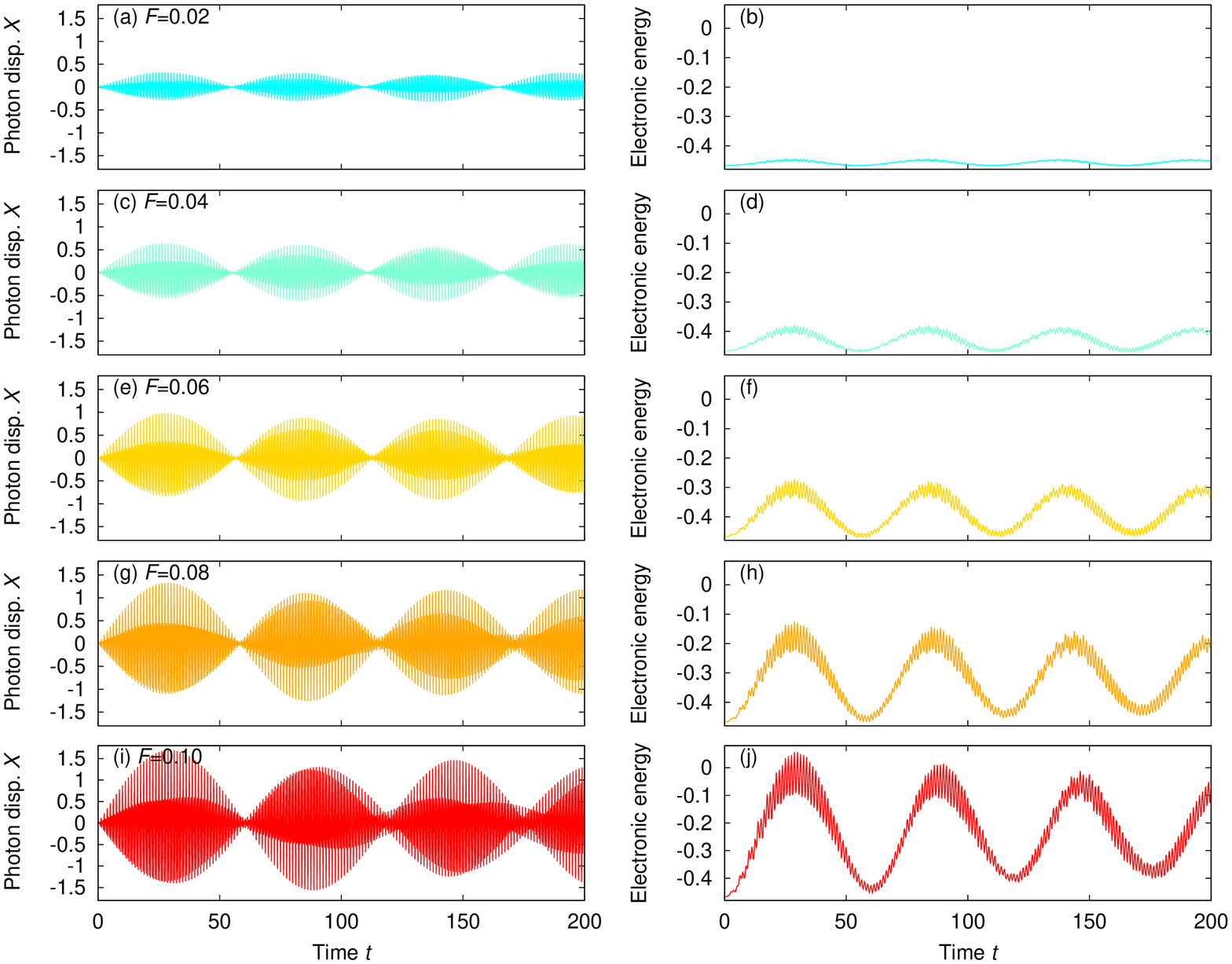}
\caption{Photon displacement and electronic energy in the high-frequency driven system at $\Omega/t_{h}$ $=$ $\Omega_{\text{dr}}/t_{h} = 10$. (a) Photon displacement (Eq.~(\ref{eq:disp})) and (b) electronic energy (Eq.~(\ref{eq:energy})) as a function of time for field strength $F=0.02$. (c-j) The same for increasing driving field strengths $F=0.04$ $\dots$ $0.10$ as labelled. Color code corresponds to the one used in Fig.~\ref{fig:s}.}
\label{fig:amp}
\end{figure*}

\begin{figure*}[htp]
\centering
\includegraphics[width=\textwidth]{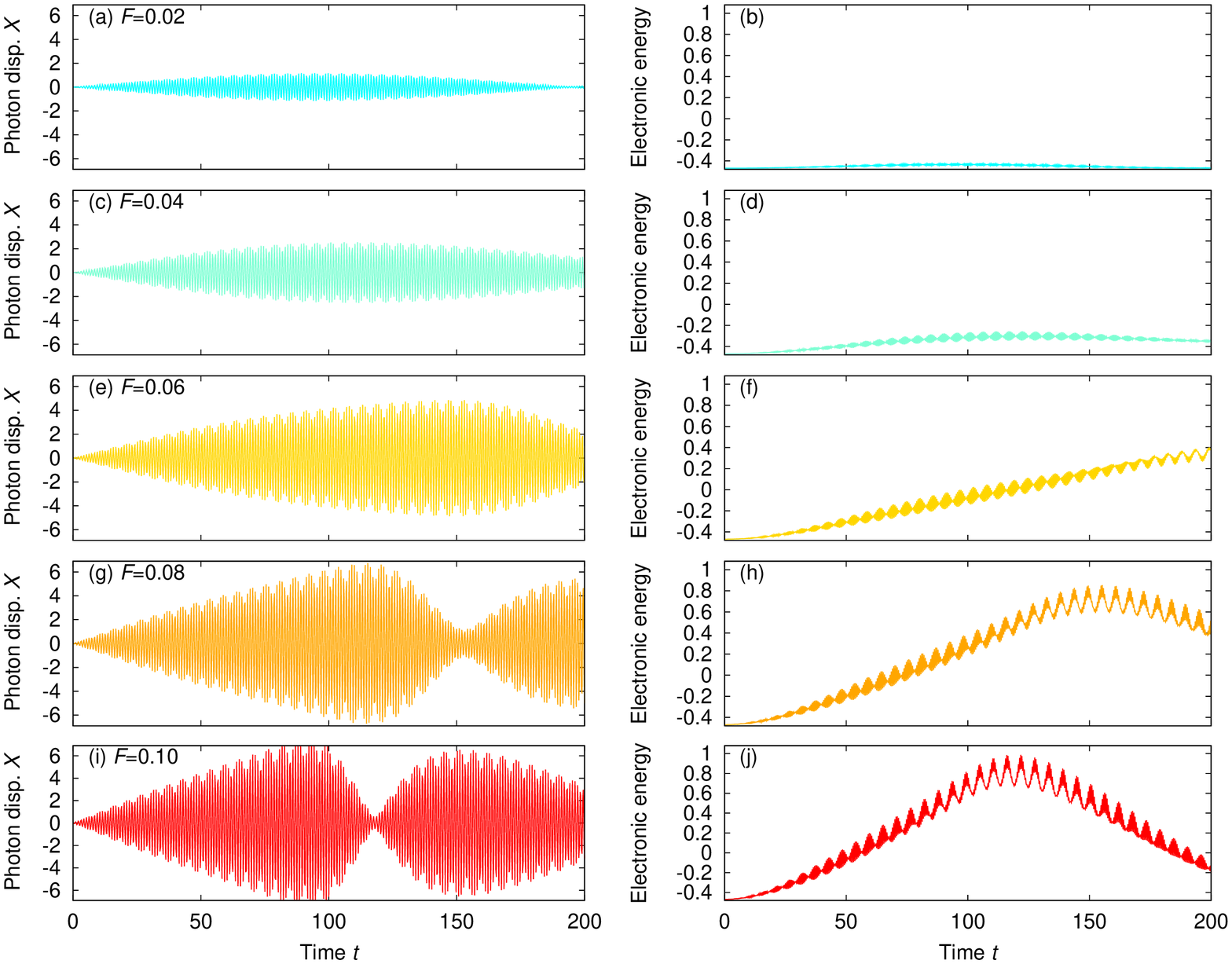}
\caption{Photon displacement and electronic energy in the sub-$U$-frequency driven system at $\Omega/t_{h}$ $=$ $\Omega_{\text{dr}}/t_{h} = 6$. (a) Photon displacement (Eq.~(\ref{eq:disp})) and (b) electronic energy (Eq.~(\ref{eq:energy})) as a function of time for field strength $F=0.02$. (c-j) The same for increasing driving field strengths $F=0.04$ $\dots$ $0.10$ as labelled. Color code corresponds to the one used in Fig.~\ref{fig:s_sub}.}
\label{fig:amp_sub}
\end{figure*}

In order to understand the behavior of the cavity-dimer system under classical driving, we show in Figs.~\ref{fig:amp} and \ref{fig:amp_sub} the time evolution of the photon displacement field
\begin{align}
    X(t) &= \langle \psi(t) | \hat{a}^{} + \hat{a}^{\dagger} | \psi(t) \rangle,
    \label{eq:disp}
\end{align}
defined for simplicity without a usually included factor of $1/\sqrt2$,
and the electronic energy (including the light-matter coupling phase terms) 
\begin{align}
    E_{\text{el}}(t) &= \langle \psi(t) | \hat{H}_{\text{el}} | \psi(t) \rangle,
    \label{eq:energy}
\end{align}
which become time-dependent through the time-dependent wave function in the driven system. 

First, for the blue-detuned case (Fig.~\ref{fig:amp}) the amplitude of the photon displacement (left panels) increases roughly linearly with the external field strength. A beating pattern is found on top of the fast oscillation with the external field, corresponding to a slight splitting (10.00 versus 10.13, see Appendix \ref{app:spectra}, Fig.~\ref{fig:fft}) of the frequencies in the photon response due to light-matter coupling. This splitting indicates the emergence of a small energy scale 0.13 in the driven system, which in fact is observed as a small shoulder developing to the left of the main peak (not shown on the scale of Fig.~\ref{fig:s}). At the same time, energy is absorbed by the electrons periodically with the beating frequency but overall the heating remains under control here (Fig.~\ref{fig:amp}, right panels). 

Next, for the red-detuned case (Fig.~\ref{fig:amp_sub}) the amplitude of the photon displacement (left panels) is overall much larger than for the blue-detuned case. Again, a beating pattern emerges, but this time the splitting of energies depends itself on the driving field strength (see Appendix \ref{app:spectra}), increasing for stronger driving resulting in shorter periods of beating. A splitting of up to 0.06 for the largest $F=0.10$ is found. Since the overall amplitude is larger here, one can clearly see multiple sidepeaks (lowest curve in Fig.~\ref{fig:s_sub}) split from the main peaks by 0.06. Thus the sidepeak emergence in the dynamical spin susceptibility of the driven system can be explained by the dynamical behavior of the driven photon-matter system. Essentially, additional spin exchange channels open up in the driven system, in which the electrons can tunnel while inelastically emitting energy into the driven cavity.  
\label{app:spectra}
\begin{figure}[htp]
\centering
\includegraphics[width=0.48\textwidth]{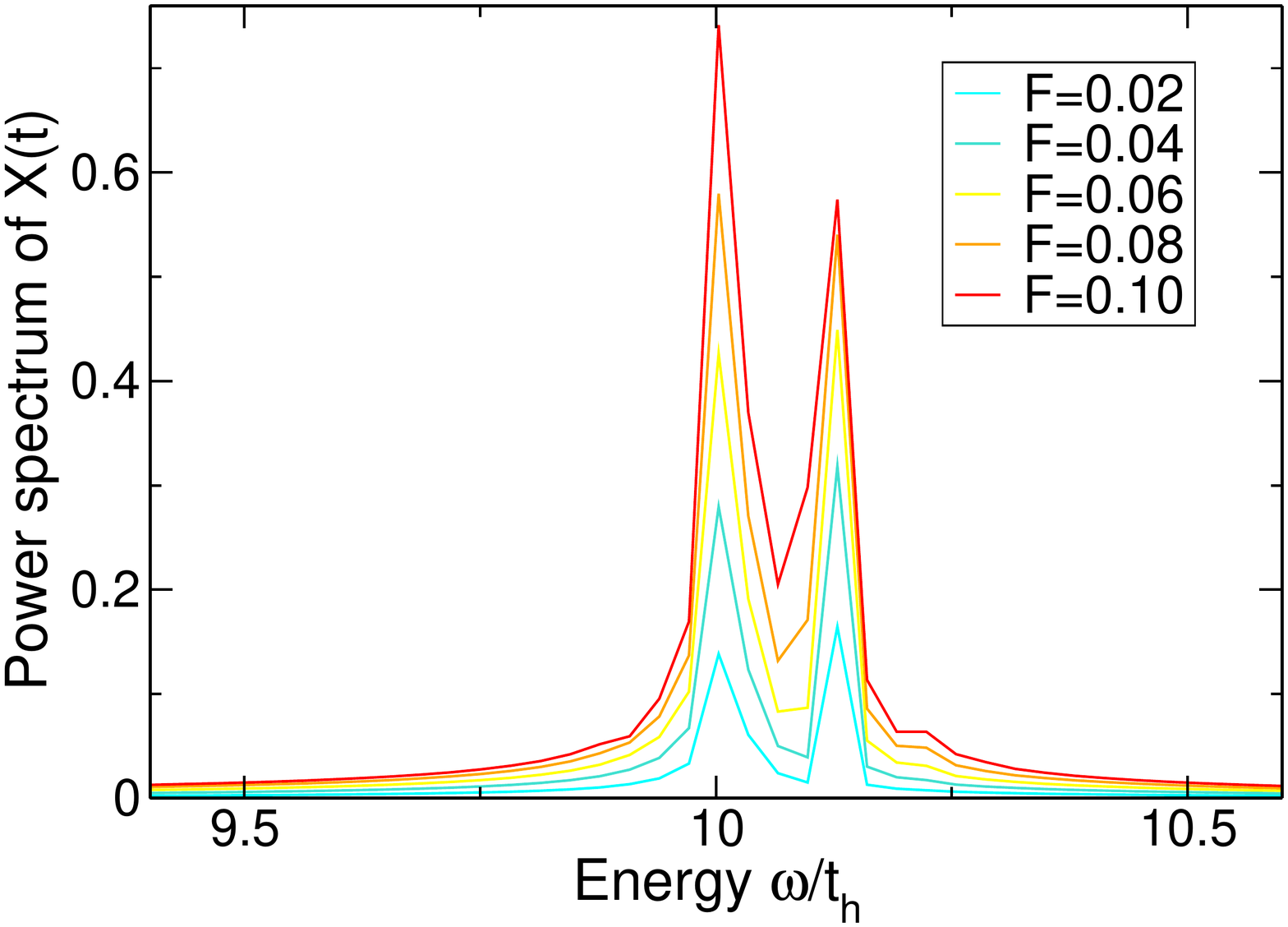}~\includegraphics[width=0.48\textwidth]{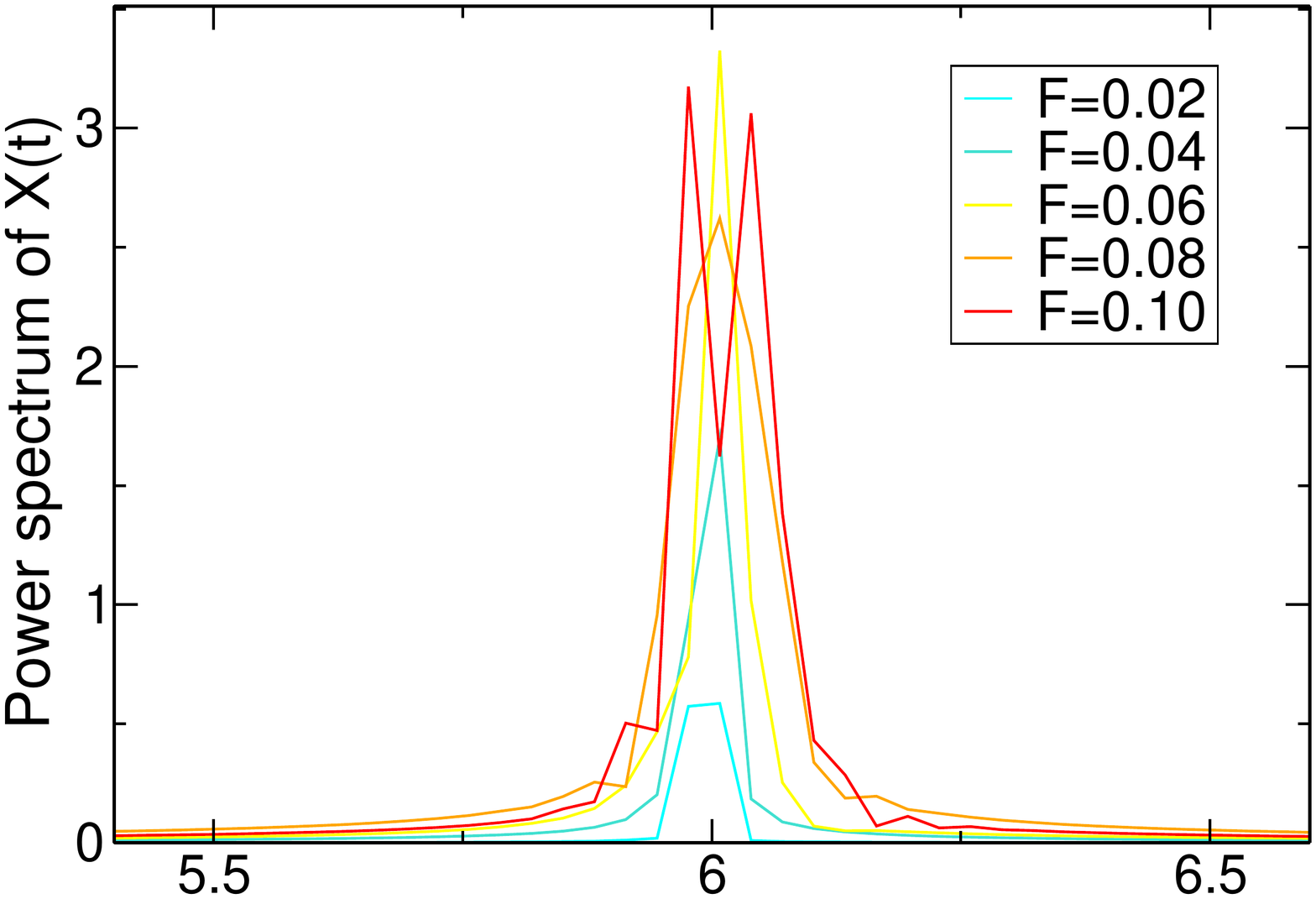}
\caption{Left: Power spectrum Fourier analysis of photon displacements shown in Fig.~\ref{fig:amp} for the high-frequency, blue-detuned case. 
Right:Power spectrum Fourier analysis of photon displacements shown in Fig.~\ref{fig:amp_sub} for the sub-$U$-frequency, red-detuned case. 
}
\label{fig:fft}
\end{figure}

Figs.~\ref{fig:fft} and \ref{fig:fft_sub} show the power spectra corresponding to the photon displacements shown in Fig.~\ref{fig:amp} and \ref{fig:amp_sub}. The beating patterns observed in the real-time data show up as peak splittings in the Fourier spectra, which is consistent with a squeezing and frequency shift of the photon mode.

\end{document}